\documentclass[letterpaper]{article} 
\usepackage{aaai2026}  
\usepackage{times}  
\usepackage{helvet}  
\usepackage{courier}  
\usepackage[hyphens]{url}  
\usepackage{graphicx} 
\usepackage{natbib}  
\usepackage{caption} 
\usepackage{algorithm}
\usepackage{algorithmic}
\usepackage{textcomp}
\usepackage{url}
\usepackage{verbatim}
\usepackage{graphicx}
\usepackage{cite}
\usepackage{multirow}
\usepackage{amssymb}
\usepackage{amsmath}
\usepackage{subfig}
 \usepackage{booktabs} 
\usepackage{newfloat}
\usepackage{listings}
\DeclareCaptionStyle{ruled}{labelfont=normalfont,labelsep=colon,strut=off} 
\floatstyle{ruled}
\newfloat{listing}{tb}{lst}{}
\floatname{listing}{Listing}
\title{TMDC: A Two-Stage Modality Denoising and Complementation Framework for Multimodal Sentiment Analysis with Missing and Noisy Modalities}
\author{
    Yan Zhuang\equalcontrib \textsuperscript{\rm 1}, Minhao Liu\equalcontrib \textsuperscript{\rm 1,2}, Yanru Zhang \textsuperscript{\rm 1,2}, Jiawen Deng\textsuperscript{\rm 1}\thanks{Corresponding authors.}, Fuji Ren\textsuperscript{\rm 1,2}\footnotemark[2]
}
\affiliations{
    \textsuperscript{\rm 1}University of Electronic Science and Technology of China\\
    \textsuperscript{\rm 2}Shenzhen Institute for Advanced Study, UESTC\\
    202211081370@std.uestc.edu.cn, \{minhaoliu,yanruzhang,dengjw,renfuji\}@uestc.edu.cn
}

\begin{document}

\maketitle

\begin{abstract}
Multimodal Sentiment Analysis (MSA) aims to infer human sentiment by integrating information from multiple modalities such as text, audio, and video. In real-world scenarios, however, the presence of missing modalities and noisy signals significantly hinders the robustness and accuracy of existing models. While prior works have made progress on these issues, they are typically addressed in isolation, limiting overall effectiveness in practical settings. To jointly mitigate the challenges posed by missing and noisy modalities, we propose a framework called \textbf{Two-stage Modality Denoising and Complementation (TMDC)}. TMDC comprises two sequential training stages. In the Intra-Modality Denoising Stage, denoised modality-specific and modality-shared representations are extracted from complete data using dedicated denoising modules, reducing the impact of noise and enhancing representational robustness. In the Inter-Modality Complementation Stage, these representations are leveraged to compensate for missing modalities, thereby enriching the available information and further improving robustness. Extensive evaluations on MOSI, MOSEI, and IEMOCAP demonstrate that TMDC consistently achieves superior performance compared to existing methods, establishing new state-of-the-art results.
\end{abstract}

\section{Introduction}
Multimodal Sentiment Analysis (MSA) leverages data from multiple modalities, such as text, video, and audio, to predict the emotional state \cite{zadeh2016mosi}. With the rapid advancements in multimedia and multimodal learning  \cite{zadeh2018multimodal,liang2022foundations}, more and more researchers are focusing on MSA. However, practical applications often face two key challenges: missing modalities, which is caused by privacy concerns \cite{jaiswal2020privacy, zhao2021missing} or incomplete data collection \cite{liu2021face}, and noisy inputs from real-world sensors. These factors severely hinder the reliability and effectiveness of MSA systems. 

\begin{figure}
    \centering
    \includegraphics[width=0.94\linewidth]{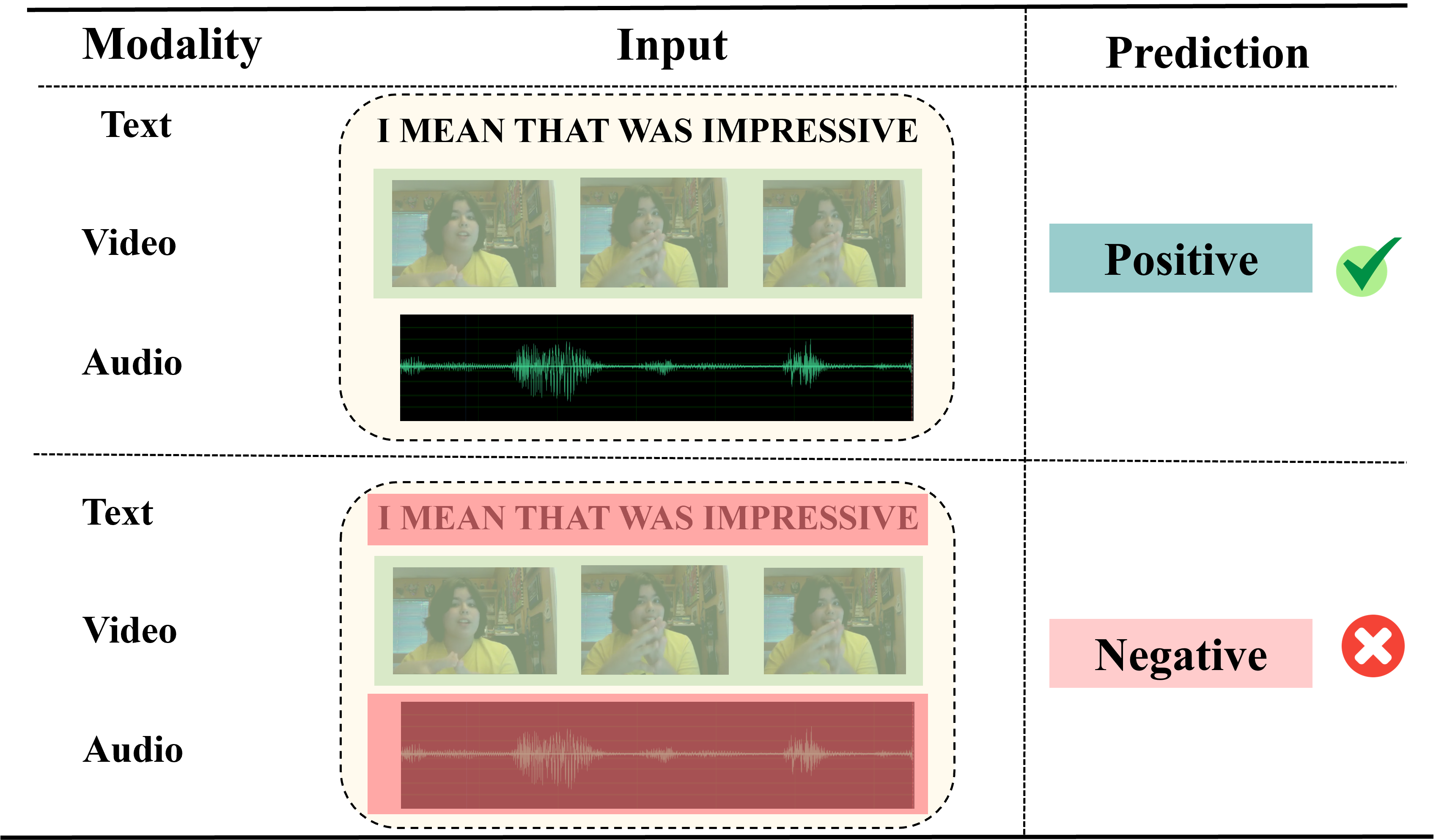}
    \caption{The existing model yields correct predictions when the input contains only missing modalities (highlighted in green), but fails when both missing (green) and noisy modalities (red) are present.}
    \label{fig:motivation}
\end{figure}

To tackle these challenges, most existing studies treat missing and noisy modalities as separate problems and develop isolated solutions. For noisy inputs, many works adopt information-theoretic approaches, such as the use of the information bottleneck \cite{mai2022multimodal}, to suppress irrelevant noise. On the other hand, methods targeting missing modalities often design reconstruction mechanisms to restore missing signals from available ones. For example, MPLMM \cite{guo2024multimodal} distills knowledge from pre-trained models into learnable prompts to compensate for missing data. IMDer \cite{wang2024incomplete} trains diffusion models on complete datasets and applies them at inference time to reconstruct absent modalities. DiCMoR \cite{wang2023distribution} creates category-specific flow-based generators for restoration, while MoMKE \cite{xu2024leveraging} trains multiple modality-specific experts and combines their outputs to form joint representations. Despite their advances, these methods typically assume clean inputs and overlook the impact of noisy data. As illustrated in Figure \ref{fig:motivation}, they often perform well under missing-modality conditions but fail when noise and missing data co-occur. Errors from noisy inputs or inaccurate reconstruction compound during training and inference, ultimately harming overall performance.

To jointly address both challenges, we propose a Two-stage Modality Denoising and Complementation (TMDC) framework. TMDC adopts a two-stage training paradigm. In the Intra-Modality Denoising Stage, TMDC is trained on complete data to capture both denoised modality-specific and modality-common representations for each modality. To handle noise, TMDC includes two denoising modules. A modality-specific denoising module, composed of a Variational Information Bottleneck (VIB) \cite{alemi2022deep}, attention layers \cite{vaswani2017attention}, and fully connected networks, filters noise while preserving distinctive features of each modality. A modality-common module learns to extract noise-robust features shared across modalities. In the Inter-Modality Complementation Stage, the denoised outputs from available modalities are used to complement missing ones by leveraging both shared and specific information extracted during the first stage. The final emotion prediction is made by integrating all observed and reconstructed representations through a fully connected layer.

The key contributions of this paper are as follows:
\begin{itemize}
\item TMDC, a framework that simultaneously addresses both missing and noisy modalities is proposed.
\item TMDC complements missing modalities using both modality-invariant and modality-specific information from available modalities.
\item Extensive experiments on multiple datasets, including scenarios with varying levels of data noise, demonstrates that TMDC achieves state-of-the-art performance.
\end{itemize}

\section{Related Work}
\subsection{Multimodal Sentiment Analysis}
Multimodal Sentiment Analysis (MSA) integrates text, video, and audio signals to predict sentiment state\cite{zadeh2016mosi}. Most existing methods assume the availability of all three modalities and focus on designing sophisticated fusion networks to integrate these heterogeneous sources effectively \cite{zong2023acformer,hazarika2020misa,lin2023dynamically, wang2022cross, mai2023multimodal, zeng2023multimodal, zhu2024kebr,yan2025r3dg}. Some approaches prioritize modality alignment; for instance, MISA \cite{hazarika2020misa} and FactorCL \cite{liang2024factorized} map each modality into modality-shared and modality-specific representations, or task-relevant and task-irrelevant components. Similarly, MULT \cite{tsai2019multimodal} and AcFormer \cite{zong2023acformer} employ cross-modal attention mechanisms to align pairs of modalities. Other methods refine fusion at a more granular level. GLoMo \cite{zhuang2024glomo} utilizes MoE networks to extract fine-grained local information for enhanced integration, while PS2RI \cite{fang2024sentiment} incorporates sarcasm-aware cues to aid sentiment prediction. KEBR \cite{zhu2024kebr}, on the other side, explores common sentimental knowledge in unlabeled videos to enrich representation. Although these models perform well when all modalities are available, their effectiveness deteriorates significantly when one or more modalities are missing, a common scenario in real-world applications, which limits their practical usability \cite{guo2024multimodal,li2024correlation, li2024unified,wei2023mmanet}.

\subsection{Incomplete Multimodal Learning}
With the growing demand for robust multimodal models, researchers have increasingly focused on incomplete multimodal learning. Most existing methods attempt to reconstruct the missing modality using pre-trained external knowledge. For example, MMIN \cite{zhao2021missing} involves pre-training on complete datasets before transferring or fine-tuning in situations with missing modalities. MPLMM \cite{guo2024multimodal} leverages external datasets and distills cross-modal information into learnable prompts to supplement missing data. IMDer \cite{wang2024incomplete} train diffusion models on complete datasets for each modality, later using them for reconstruction when a modality is missing. Similarly, DiCMoR \cite{wang2023distribution} employs category-specific flow-based models to restore absent modalities. MoMKE \cite{xu2024leveraging} takes a different approach by training modality-specific MoE networks on complete datasets to capture modality-specific representations and then fusing them into a joint representation.  

While these methods have achieved notable success, they assume that the collected data is noise-free. However, real-world multimodal data often contains inherent noise \cite{mai2022multimodal, gao2024embracing}, and errors introduced during reconstruction further degrade performance. This reduces the robustness of these models, particularly in noisy environments. To address this limitation, we propose TMDC, a framework that explicitly considers both intrinsic noise and errors introduced by missing modalities. By employing a denoising-first learning approach, TMDC enhances representation robustness, making it more suitable for real-world applications.
\section{Methodology}

\subsection{Problem Definition}
Given a dataset $D=\{X^A_i,X^T_i,X^V_i\}_{i=1}^N$ comprising three modalities (e.g. text (`T'), video (`V'), and audio (`A')), each element $X^m_i$ represents the representation of modality $m$ for $i^{th}$ instance, $m\in\{A,T,V\}$. Specifically, $X^m_i\in\mathcal{R}^{L_m\times D_m}$, where $L_m$ denotes the sequence length, and $D_m$ represents the feature dimension of modality $m$. For simplicity, we omit the subscript $i$ and use $X^m$ to denote the representation of modality $m$. In real-world scenarios, some modalities may be missing due to various factors. To denote missing modalities, we use $\hat{X^m}$. Additionally, modality representations inherently contain noise, which we do not explicitly annotate. The goal of MSA with missing and noisy modalities is to train effective and robust models under different missing-modality and noisy-modality conditions.

\begin{figure*}[ht]
\centering 
\includegraphics[width=1.6\columnwidth]{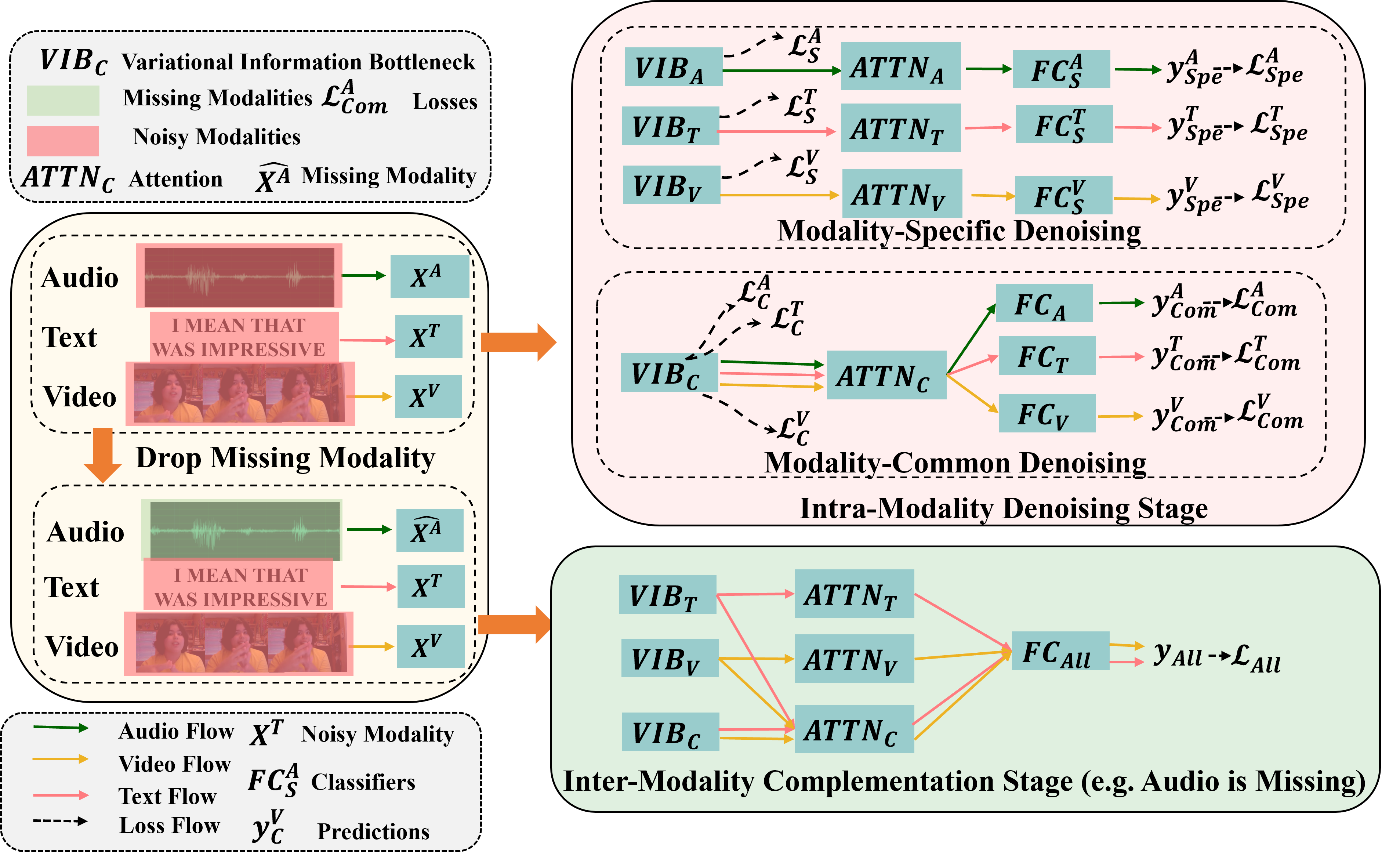}
\caption{Illustration of the proposed TMDC. TMDC includes two training stages. In the first stage, TMDC learns from complete modality information using two denoising modules. The modality-specific denoising module applies separate networks to each modality to remove noise while preserving unique modality information. Simultaneously, the modality-common denoising module employs a shared network to filter noise across multiple modalities and extract common information. In the second stage, the learned shared information is used to supplement missing modalities.}\label{framework}
\end{figure*}

\subsection{Preliminaries}
Given a modality representation $X^m$ with inherent noise, VIB approximates the information bottleneck by learning a compressed encoding $X^m_s$ that retains essential task-relevant information while filtering out unnecessary information. The optimization objective is formulated as:
\begin{equation}
\mathcal{L}^{m}=I(X^m_s,y)-\beta I(X^m_s,X^m).\label{eq1}
\end{equation}
Here $y$ is the ground truth, $I(\cdot,\cdot)$ denotes mutual information, measuring the correlation between representations. A higher mutual information value indicates stronger relevance. $X^m_s$ represents the denoised modality representation, and $\beta$ is a Lagrange multiplier.

The mutual information $I(X^m_s,y)$ can be implemented using task-related loss $\mathcal{L}_{TASK}(y^m,y)$ and  $I(X^m_s,X^m)$ can be approximated by employing KL divergence, leading to the revised objective:
\begin{equation}
\mathcal{L}^{m}=\mathcal{L}_{TASK}(y^m,y)+\beta KL(p(e^m_s|e^m)||\mathcal{N}(0,\textbf{I})).\label{eq2}
\end{equation}
Here $e_s^m\in X_s^m$, $e^m\in X^m$, $p(e^m_s|e^m)\sim \mathcal{N}(\mu^m_s, (\sigma^m_s)^2\textbf{I})$, and $\mu^m_s$, $\sigma^m_s$ denote the mean and variance of the denoised representation, which are predicted using fully connected layers through:
\begin{equation}
\mu^m_s=W^m_{1}e^m+b_{1}^m,\label{eq3}
\end{equation}
and:
\begin{equation}
\sigma^m_s=W^m_{2}e^m+b_{2}^m.\label{eq4}
\end{equation}
Following the reparameterization trick \cite{kingma2013auto} as in \cite{mai2022multimodal, gao2024embracing, alemi2022deep}, denoised representation can be constructed as:
\begin{equation}
X^m_s=\mu^m_s+\epsilon \sigma^m_s.\label{eq5}
\end{equation}
Here $\epsilon\in \mathcal{N}(0,\textbf{I})$. $y^m$ is the predicted label using denoised representation $X^m_s$ through a linear layer:
\begin{equation}
y_{s}^m=W^m_{3}X^m_s+b_{3}^m.\label{eq6}
\end{equation}
Here $W^m_1$, $W^m_2$, $W^m_3$ are trainable weights, and $b^m_1$, $b^m_2$, $b^m_3$ are bias terms. 

\subsection{Model Overview}
We propose TMDC, a two-stage modality denoising and complementation framework to handle missing and noisy modalities in MSA. As illustrated in Figure \ref{framework}, TMDC consists of the Intra-Modality Denoising (IMD) Stage and the Inter-Modality Complementation (IMC) Stage. In the IMD Stage, TMDC learns denoised modality-specific and modality-invariant representations from complete multimodal data. The Modality-Specific Denoising (MSD) Module, which incorporates modality-specific VIB \cite{alemi2022deep} and attention layers \cite{vaswani2017attention}, extracts modality-specific representations. Meanwhile, the Modality-Common Denoising (MCD) Module, utilizing shared VIB and attention layers, captures shared representations across modalities. While in the IMC Stage, TMDC processes incomplete data with missing modalities. The available modality-specific and modality-invariant representations are leveraged to complement missing information. The final integrated representation is then passed through a fully connected layer for sentiment prediction. The following sections provide a detailed introduction of each component and process within TMDC.

\subsection{Intra-Modality Denoising Stage}
This section details the first stage of TMDC. Following prior works \cite{wang2024incomplete, wang2023distribution, guo2024multimodal, xu2024leveraging}, TMDC is initially trained on the complete dataset to obtain comprehensive information from all modalities. However, unlike existing approaches that overlook inherent noise during training, TMDC explicitly addresses this issue by introducing two modules: the Modality-Specific Denoising (MSD) Module and the Modality-Common Denoising (MCD) Module. These modules separately extract modality-specific and shared information while filtering out noise. Below, we describe each module in detail. 

Since different modalities have different dimensions, following existing studies\cite{tsai2019multimodal}, we use the 1D temporal convolutional layer (Conv1D) with a kernel size of $3\times 3$ to standardize each modality to the same dimension ($D$) and sequence length ($T$) through:
\begin{equation}
X^m=W^m_{3\times 3}(X^m).\label{eq0}
\end{equation}
Here $X^m\in\mathcal{R}^{T\times D}$, and $W^m_{3\times 3}$ is the trainable weights.

\subsubsection{Modality-Specific Denoising Module}
This module aims to reduce noise within each modality and extract modality-specific representations. To achieve this, TMDC employs a distinct yet structurally identical network for each modality. Inspired by prior methods \cite{mai2022multimodal, gao2024embracing}, we adopt VIB \cite{alemi2022deep} to simultaneously remove noise and redundant information, which is introduced in Preliminaries Section. Specifically, for each modality $m$, we use the VIB to obtain denoised representations $X_s^m$ and get the predicted label $y_s^m$ using Equations \ref{eq3}-\ref{eq6}. Once the denoised representations are obtained, they are processed through a modality-specific attention network, which consists of a multi-head attention (MHA) layer and a residual fully connected layer, enabling interactions among different modality representations through:
\begin{equation}
X^m_{Spe}=MHA^m(X^m_s,X^m_s)+X^m_s,\label{eq7}
\end{equation}
and:
\begin{equation}
\hat{X}^m_{Spe}=X^m_{Spe}+W^m_{4}X^m_{Spe}+b^m_{4}.\label{eq8}
\end{equation}
To ensure these representations retain modality-specific information, a fully connected layer predicts the label through:
\begin{equation}
y^m_{Spe}=W^m_5\hat{X}^m_{Spe}+b^m_{5}.\label{eq9}
\end{equation}
Notably, each modality has its own independent network, as illustrated in Figure \ref{framework}, and the network parameters are not shared across modalities.

\subsubsection{Modality-Common Denoising Module}
This module extracts denoised, modality-invariant features shared across all modalities. Its architecture is similar to the MSD Module, with a key difference: the parameters in Conv1D, VIB and Attention layers are shared, as shown in Figure \ref{framework}.

Specifically, denoised modality-invariant representation $X^m_c$ for each modality $m$ is obtained using Equations \ref{eq3}-\ref{eq5}. And the representation after attention interaction is obtained, denoted as $\hat{X}^m_{Com}$ for modality $m$, by changing query and key to $X_c^m$ in Equation \ref{eq7} and changing $X_{Spe}^m$ to $X_{Com}^m$ in Equation \ref{eq8}. The predicted label $y_c^m$ from the denoised modality-invariant representation $X_c^m$, and $y_{Com}^m$ from the interaction representation $\hat{X}^m_{Com}$ for each modality $m$ are then obtained through separate linear layers.

\subsection{Inter-Modality Complementation Stage}
To simulate real-world conditions where certain modalities may be unavailable, we randomly set the representation of the missing modality to a zero vector during training. Without loss of generality, we assume the audio modality (A) is missing in this case (as illustrated in Figure \ref{framework}), resulting in model input of the form $\{\hat{X}^A, X^T, X^V\}$.

To compensate for the missing information, we leverage the available modalities through a structured fusion strategy. For the text and video modalities, we first extract two types of representations using VIB: (1) modality-specific representations $X^T_s$ and $X^V_s$; and (2) modality-invariant representations $X^T_c$ and $X^V_c$ using Equations \ref{eq3}-\ref{eq5}. To enhance the expressiveness of uni-modal representations, we integrate both type of representations with modality-specific attention layer trained in the first stage, where $X^{m1}_s$ serves as the query, and $X^{m1}_c$ acts as the key and value through:
\begin{equation}
X^{m1}_{All}=MHA^{m1}(X_s^{m1}, X_c^{m1}),\label{eq14}
\end{equation}
and:
\begin{equation}
\hat{X}_{All}^{m1}=X^{m1}_{All}+W^{m1}_{All}X^{m1}_{All}+b^{m1}_{All}.\label{eq15}
\end{equation}
Here $m1\in\{T,V\}$. 

To further compensate for the missing modality, we model cross-modal dependencies between the available modalities using a bidirectional attention mechanism. Specifically, we extract complementary features by swapping the roles of the query and key in the attention layers in Equation \ref{eq14}: $X_{T2V}$ is obtained by setting the query to $X_c^T$ and the key/value to $X_s^V$, and vice versa for $X_{V2T}$. The corresponding enhanced features $\hat{X}_{T2V}$ and $\hat{X}_{V2T}$ are computed using the same transformation as in Equation \ref{eq15}.

We then fuse the cross-modal features by adding $\hat{X}_{T2V}$ and $\hat{X}_{V2T}$ to produce a compensated representation $X_{Compensate}$. The final multimodal representation is constructed by concatenating the compensated features with the refined text and video representations, followed by a fully connected layer for sentiment prediction:
\begin{equation}
X = [X_{Compensate}, \hat{X}_{All}^T, \hat{X}_{All}^V],\label{eq21}
\end{equation}
\begin{equation}
y_{All} = W_{All}X + b_{All}.\label{eq22}
\end{equation}

In cases where multiple modalities are missing, for example, when only the audio modality is available, a different compensation strategy is applied. The available modality undergoes the same refinement as described above. For the missing modalities, we approximate their representations using self-attention within the available modality by changing query and key to $X_c^A$ and $X_s^A$ in Equation \ref{eq14} and transform the obtained representation to get $\hat{X}_{A2A}$ through Equation \ref{eq15}. Finally, the multimodal representation in this scenario is formed by repeating the compensated audio representation:
\begin{equation}
X = [\hat{X}^A_{All}, \hat{X}_{A2A}, \hat{X}_{A2A}].\label{eq25}
\end{equation}

\subsection{Training Objectives}\label{training_loss}
This section outlines the training objectives for both IMD and IMC stages in TMDC framework. Since MSA involves both classification and regression tasks, we first introduce the task-specific loss function, denoted as $\mathcal{L}_{TASK}$. For regression tasks, we use Mean Squared Error (MSE) loss, while for classification tasks, the Cross-Entropy loss is applied.

In the IMD Stage, TMDC is trained on the complete dataset to learn both modality-specific and modality-invariant representations. The training objective includes two components: (1) VIB loss using Equation \ref{eq2} to encourage denoised representation learning; and (2) task-specific losses applied to representations after modality interactions. In order to distinguish different task-specific losses, we define the loss of different representations as:
\begin{equation}
\mathcal{L}_b^m=\mathcal{L}_{TASK}(y_b^m,y). 
\end{equation}
Here $b\in\{Spe,Com\}$ denotes the predictions, and $m$ denotes the modality. The overall loss function for IMD stage is formulated as:
\begin{equation}
\mathcal{L}_{IMD}=\sum_{m\in\{A,T,V\}}(\sum_{b\in\{Spe,Com\}}\mathcal{L}_b^m+\sum_{k\in\{s,c\}}\mathcal{L}_k^m).\label{26}
\end{equation}
Here $k$ denotes the loss from the modality-specific (s) or modality-invariant (c) representations using Equation \ref{eq2} in MSD or MCD module.

In the IMC Stage, TMDC is optimized based only on the final fused representation. Here, the training objective focuses solely on the prediction loss for the concatenated multimodal representation:
\begin{equation}
\mathcal{L}_{IMC}=\mathcal{L}_{TASK}(y_{All},y).\label{27}
\end{equation}

\begin{table*}[t]
\centering
\begin{tabular}{@{}cccccccccc}
\hline
\multirow{2.5}{*}{Models} & \multicolumn{7}{c}{\textbf{Modalities}} \\
\cmidrule(r){2-8}
& A & T & V & A,V & A,T & T,V & T,A,V \\
\hline
\hline
\multicolumn{8}{c}{\textbf{Results on MOSI}} \\ 
\hline
& ACC/F1 & ACC/F1 & ACC/F1 & ACC/F1 & ACC/F1 & ACC/F1 & ACC/F1 \\
\hline
MCTN & 56.10/54.50 & 79.10/79.20 & 55.00/54.40 & 57.50/57.40 & 81.00/81.00 & 81.10/81.20 & 81.40/81.50 \\
MMIN & 55.30/51.50 & 83.80/83.80 & 57.00/54.00 & 60.40/58.50 & 84.00/84.00 & 83.80/83.90 & 84.60/84.40 \\
GCNet & 56.10/54.50 & 83.70/83.60 & 56.10/55.70 & 62.00/61.90 & 84.50/84.40 & 84.30/84.20 & 85.20/85.10 \\
IMDer & 62.00/62.20 & 84.80/84.70 & 61.30/60.80 & 63.60/63.40 & 85.40/85.30 & 85.50/85.40 & 85.70/85.60 \\
DiCMoR & 60.50/60.80 & 84.50/84.40 & 62.20/60.20 & 64.00/63.50 & 85.50/85.50 & 85.50/85.40 & 85.70/85.60 \\
MPLMM & 62.71/\textbf{63.65} & 80.12/80.31 & 63.12/63.74 & 65.02/65.41 & 80.76/81.09 & 81.12/81.19 & - \\
MoMKE & \textbf{63.19}/58.61 & 86.59/86.52 & 63.35/63.34 & 64.04/64.66 & 87.20/87.17 & 87.04/87.00 & 87.96/87.89 \\
\hline
TMDC & 62.35/60.24 & \textbf{87.35/87.27} & \textbf{64.63/64.82} & \textbf{65.40/65.60} & \textbf{87.50/87.45} & \textbf{87.96/87.87} & \textbf{88.26/88.19} \\
\hline
\hline
\multicolumn{8}{c}{\textbf{Results on MOSEI}} \\ 
\hline
& ACC/F1 & ACC/F1 & ACC/F1 & ACC/F1 & ACC/F1 & ACC/F1 & ACC/F1 \\
\hline
MCTN & 62.70/54.50 & 82.60/82.80 & 62.60/57.10 & 63.70/62.70 & 83.50/83.30 & 83.20/83.20 & 84.20/84.20 \\
MMIN & 58.90/59.50 & 82.30/82.40 & 59.30/60.00 & 63.50/61.90 & 83.70/83.30 & 83.80/83.40 & 84.30/84.20 \\
GCNet & 60.20/60.30 & 83.00/83.20 & 61.90/61.60 & 64.10/57.20 & 84.30/84.40 & 84.30/84.40 & 85.20/85.10 \\
IMDer & 63.80/60.60 & 84.50/84.50 & 63.90/63.60 & 64.90/63.50 & 85.10/85.10 & 85.00/85.00 & 85.10/85.10 \\
DiCMoR & 62.90/60.40 & 84.20/84.30 & 63.60/63.60 & 65.20/64.40 & 85.00/84.90 & 84.90/84.90 & 85.10/85.10 \\
MPLMM & 67.33/68.71 & 79.12/79.17 & 67.29/69.40 & 68.21/69.91 & 80.45/80.43 & 80.11/80.13 & - \\
MoMKE & 72.56/71.03 & 86.46/86.43 & 70.12/70.23 & 73.34/71.82 & 86.68/86.61 & 86.79/86.69 & 87.12/87.03 \\
\hline
TMDC & \textbf{73.64/72.23} & \textbf{86.87/86.82} & \textbf{71.60/70.65} & \textbf{74.13/73.41} & \textbf{87.15/87.14} & \textbf{87.48/87.43} & \textbf{87.67/87.62} \\
\hline
\hline
\multicolumn{8}{c}{\textbf{Results on IEMOCAP}} \\ 
\hline
& WA/UA & WA/UA & WA/UA & WA/UA & WA/UA & WA/UA & WA/UA \\
\hline
MCTN & 49.75/51.62 & 62.42/63.78 & 48.92/45.73 & 56.34/55.84 & 68.34/69.46 & 67.84/68.34 & - \\
MMIN & 56.58/59.00 & 66.57/68.02 & 52.52/51.60 & 63.99/65.43 & 72.94/75.14 & 72.67/73.61 & - \\
IF-MMIN & 55.03/53.20 & 67.02/68.20 & 51.97/50.41 & 65.33/66.52 & 74.05/75.44 & 72.68/73.62 & - \\
MRAN & 55.44/57.01 & 65.31/66.42 & 53.23/49.80 & 64.70/64.46 & 73.00/74.58 & 72.11/72.24 & - \\
MoMKE & 70.32/71.38 & 77.82/78.37 & 58.60/54.70 & 68.85/67.65 & 79.89/79.53 & 77.87/77.84 & 80.13/79.99 \\
\hline
TMDC & \textbf{70.45/71.40} & \textbf{77.88/78.44} & \textbf{59.18/55.33} & \textbf{70.21/69.91} & \textbf{79.99/81.45} & \textbf{78.20/78.29} & \textbf{80.48/80.69} \\
\hline
\end{tabular}
\caption{Performance comparison under different modality combinations on three datasets. Best performance is bold.}\label{EX_MOSEI}
\end{table*}

\section{Experiment}
\subsection{Datasets and Evaluation Criteria}
We evaluate TMDC on three benchmark datasets: MOSI \cite{zadeh2016mosi}, MOSEI \cite{zadeh2018multimodal}, and IEMOCAP \cite{busso2008iemocap}. MOSI and MOSEI are regression-based sentiment datasets with scores from -3 to +3. MOSI includes 2,199 video clips, while MOSEI has 22,856 clips. Following prior work \cite{xu2024leveraging, wang2024incomplete, guo2024multimodal}, we convert scores into binary labels and report accuracy and F1-score. IEMOCAP contains 5,531 utterances across five sessions. We follow common practice \cite{xu2024leveraging} to classify four emotions: neutral, happy, sad, and angry. Performance is measured using weighted accuracy (WA) and unweighted accuracy (UA) under five-fold cross-validation. 

\subsection{Experiment Setups}
\textbf{Feature Extraction}
Following previous studies
\cite{lian2023gcnet,xu2024leveraging}, we adopt the same feature extraction methods for all three datasets. For the text modality, we use the pre-trained DeBERTa-large model \cite{he2021deberta} to obtain textual representations. For the audio modality, we extract features using the pre-trained wav2vec-large model \cite{schneider2019wav2vec}. For the video modality, we apply the MTCNN face detection algorithm \cite{zhang2016joint} followed by the pre-trained MA-Net model \cite{zhao2021learning} to obtain video representations. The final feature dimensions for the text, audio, and video modalities are 1024, 512, and 1024, respectively.

\textbf{Implementation Details}
Consistent with prior work \cite{ xu2024leveraging, guo2024multimodal}, we evaluate the performance of TMDC under fixed missing modality scenarios, where a specific modality is absent during training, validation, and testing. For example, in Table \ref{EX_MOSEI}, `A' indicates that only the audio modality is available. All experiments are implemented in PyTorch and conducted on a GTX 3090 GPU with CUDA 11.5. We use Torch version 1.12.1 for model training and Adam optimizer \cite{kingma2014adam} across all datasets. More implementation details and hyper-parameters are shown in Supplementary Material.

\subsection{Comparison with State-of-the-art Methods}
To comprehensively evaluate TMDC’s performance, we compare it with several state-of-the-art methods, including MCTN \cite{pham2019found}, MMIN \cite{zhao2021missing}, GCNet \cite{lian2023gcnet}, IMDer \cite{wang2024incomplete}, DiCMoR \cite{wang2023distribution}, MPLMM \cite{guo2024multimodal}, MoMKE \cite{xu2024leveraging}, IF-MMIN \cite{zuo2023exploiting}, and MRAN \cite{luo2023multimodal}. Table \ref{EX_MOSEI} presents the results on the MOSI, MOSEI, and IEMOCAP datasets.

Our findings show that TMDC outperforms all baseline models in most scenarios, with only one exception: when only the audio modality is available in the MOSI dataset. However, in all other MOSI settings, TMDC surpasses existing methods. Moreover, TMDC consistently achieves state-of-the-art results across all missing-modality scenarios in the MOSEI and IEMOCAP datasets. Specifically, on MOSI, TMDC achieves an average accuracy of 77.64 and an F1-score of 77.35, outperforming the second-best model, MoMKE, by 0.59 and 0.89, respectively. On MOSEI, TMDC attains 81.22 accuracy and 80.76 F1-score, improving upon the strongest baseline by 0.78 in both metrics. For IEMOCAP, TMDC achieves a WA of 73.77 and a UA of 73.64, exceeding the best competing method by 0.42 and 0.86, respectively. These results highlight TMDC’s effectiveness in learning robust representations.

\subsection{Ablation Study}\label{sec:abla}
To investigate how each module and stage contributes to TMDC's effectiveness, we conduct an ablation study by evaluating four TMDC variants: (1) `w/o IMD':  Excludes the Intra-Modality Denoising stage, meaning TMDC is not trained on complete datasets before transitioning to the IMC stage. (2) `w/o IMC': Removes Inter-Modality Complementation, where representations from available modalities are directly concatenated without additional compensation after the IMD stage. (3) `w/o MCD': Excludes the Modality-Common Denoising Module in both stages, preventing any compensation for missing information. (4) `w/o MSD': Removes the Modality-Specific Denoising Module in both stages, relying only on shared modality-invariant information for predictions.

Table \ref{EX_ablas} presents the average results of TMDC and its four variants across all seven missing-modality scenarios. We observe that removing any stage results in performance degradation, but the effects of removing modules vary across datasets. Overall, removing both denoising modules leads to performance degradation, with `w/o MSD' having a more pronounced impact than `w/o MCD'. This may be because modality-specific information is more discriminative than modality-invariant information, as the latter can be inferred from other modalities, whereas the former cannot. Consequently, losing modality-specific information results in a greater drop in performance. Notably, the `w/o IMC' variant suffers the most significant drop across all datasets. This is likely because, while the IMD stage has access to all modalities, the IMC stage encounters missing modalities without compensation, leading to substantial performance deterioration. More detailed ablation results for each missing modality are shown in Supplementary Material.

\subsection{Further Analysis}
In this section, we focus on TMDC's performance under noisy conditions, the training dynamics of different loss functions in the IMD stage, and the relationships between representations under various missing modality scenarios. 

\begin{figure}[h]
    \centering
    \subfloat[Losses in MSD Module]{
\includegraphics[width=0.46\columnwidth]{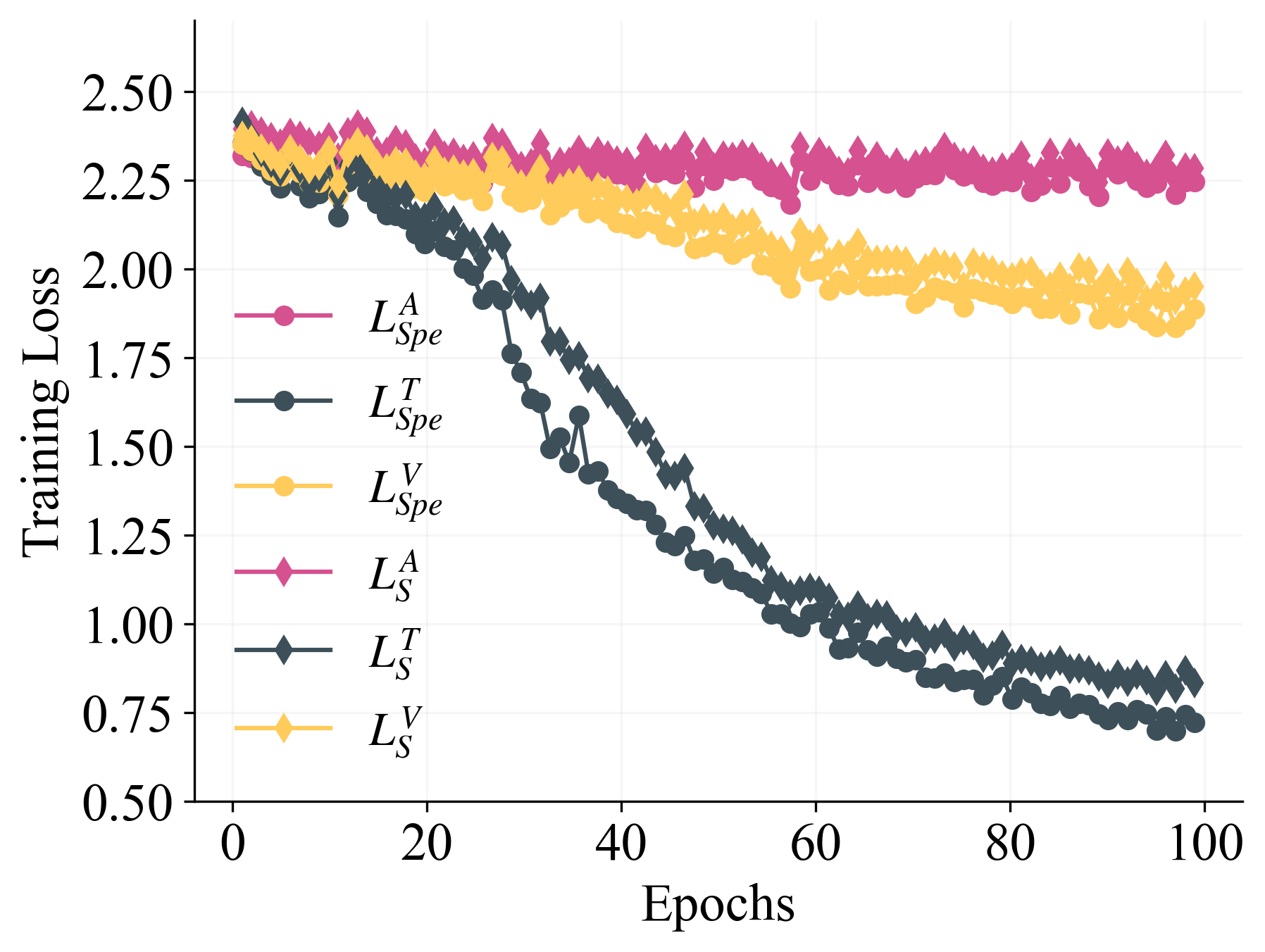}\label{loss_spe}
 }
 \subfloat[Losses in MCD Module]{
\includegraphics[width=0.46\columnwidth]{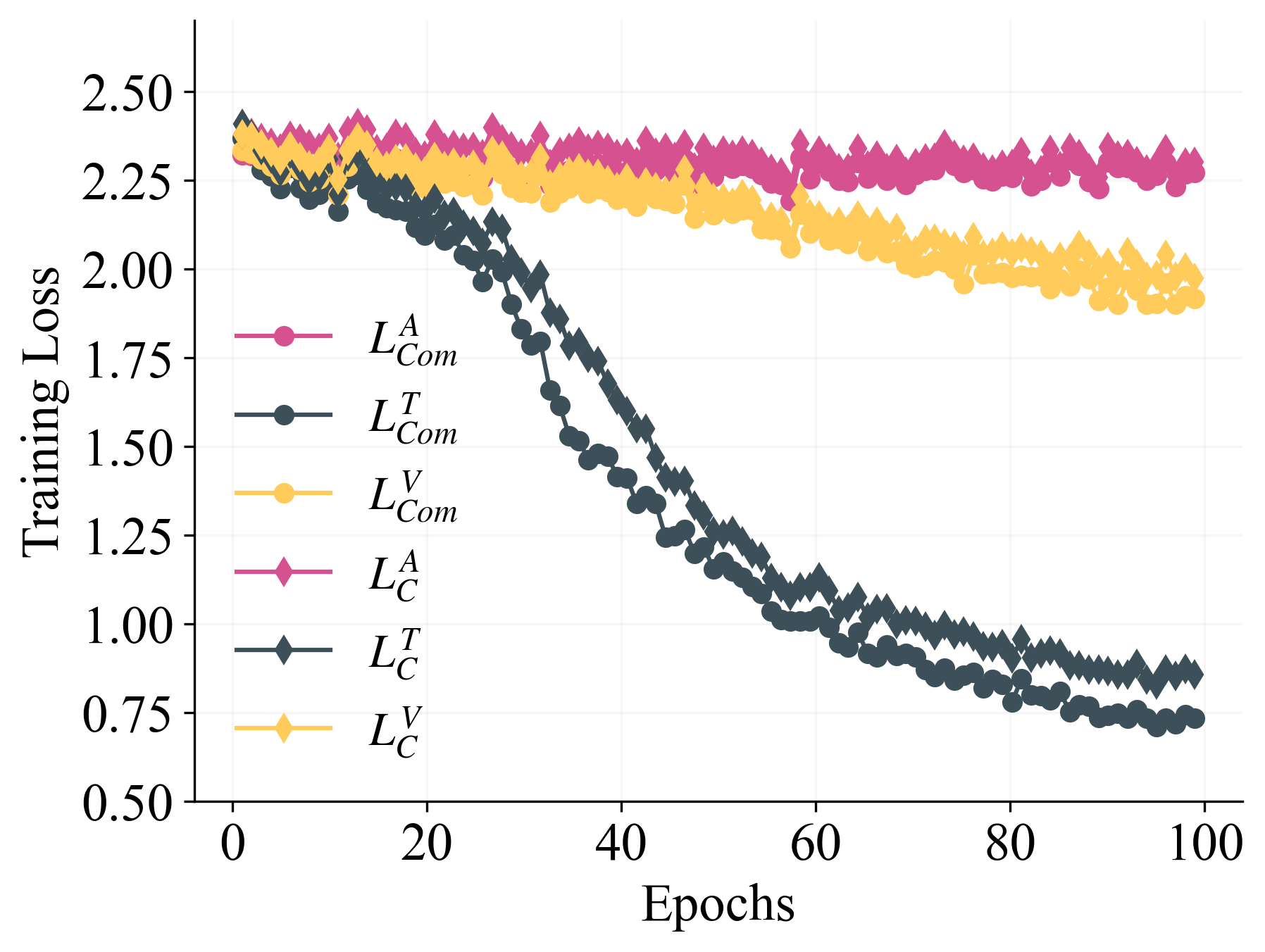}\label{loss_com}
 }
 \caption{Visualization of the training losses in IMD stage on MOSI dataset.}\label{training_losses}
\end{figure}

\begin{figure*}[h]
    \centering
    \subfloat[In IMD Stage]{
\includegraphics[width=0.48\columnwidth]{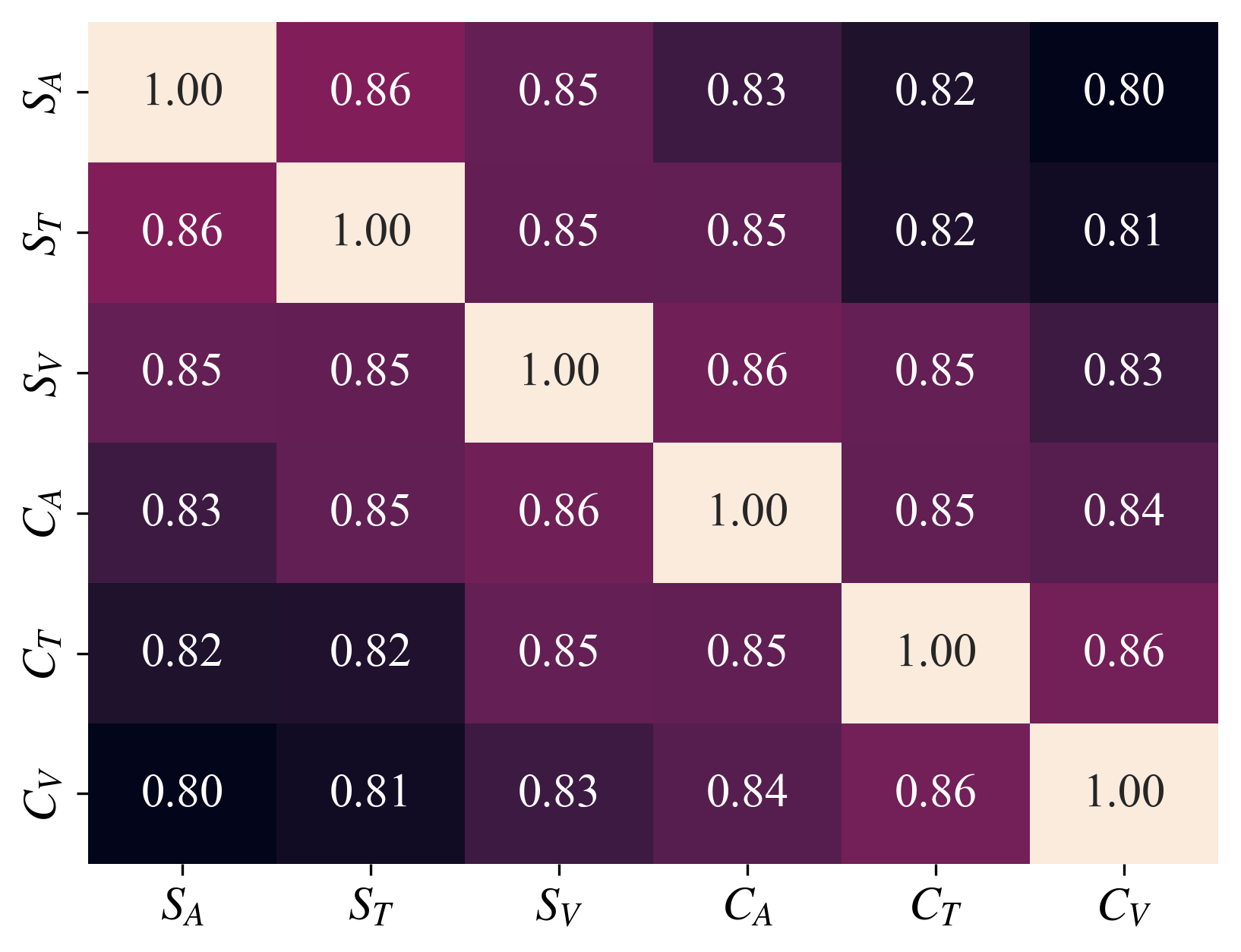}\label{first_cos}
 }
 \subfloat[In IMC Stage with `A,T,V']{
\includegraphics[width=0.48\columnwidth]{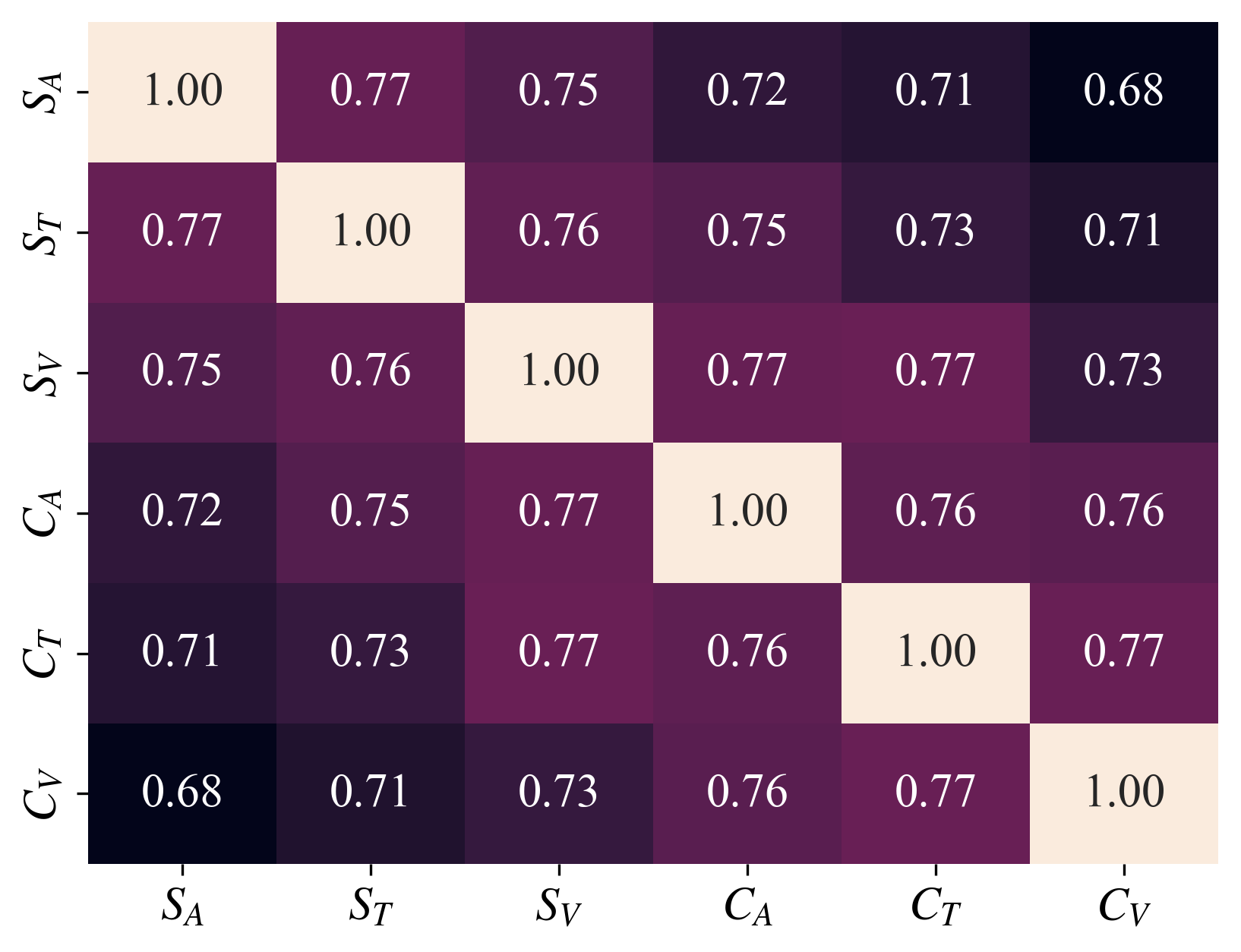}\label{second_atv_cos}
 }
  \subfloat[In IMC Stage with `A']{
\includegraphics[width=0.48\columnwidth]{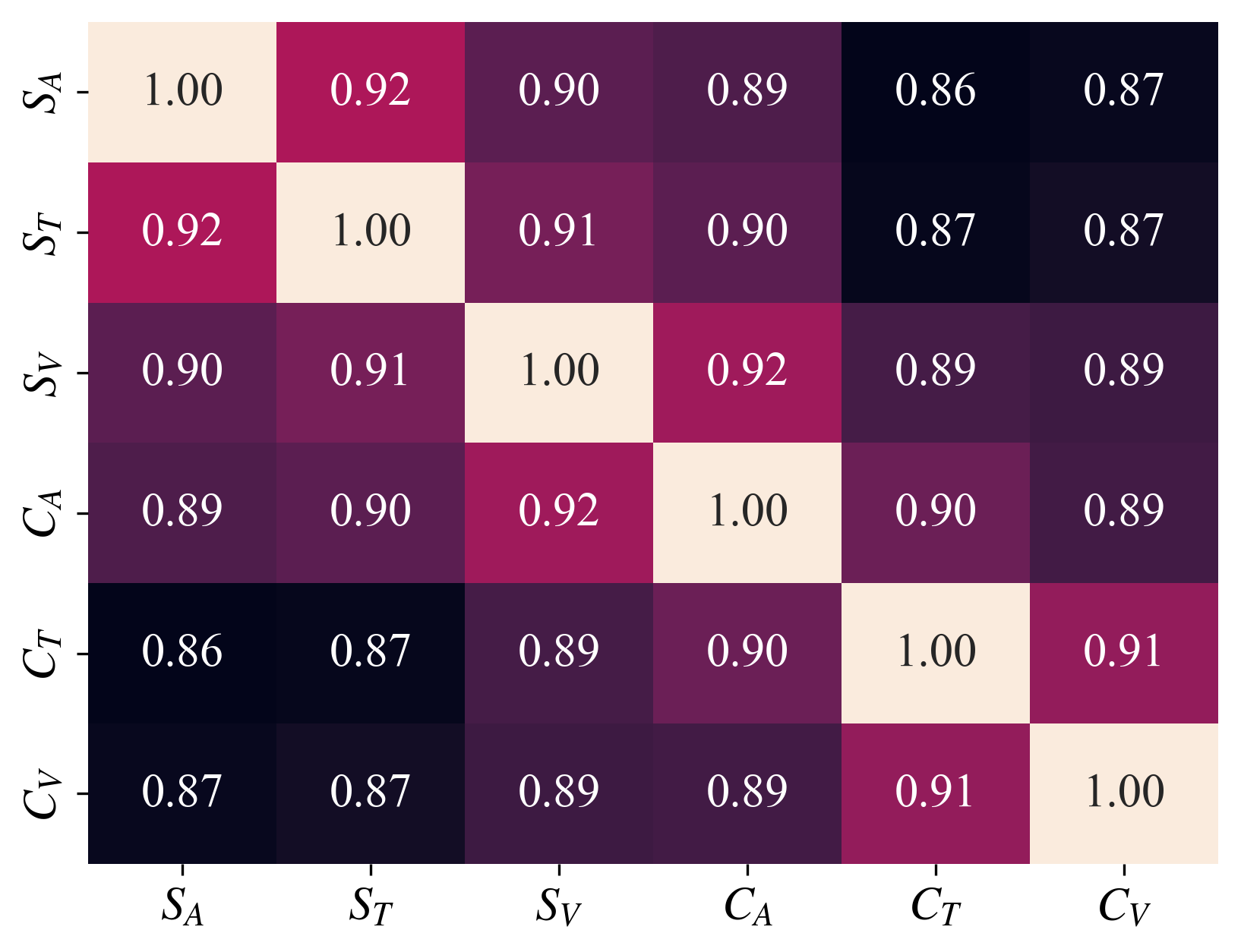}\label{second_a_cos}
 }
  \subfloat[In IMC Stage with `A,V']{
\includegraphics[width=0.48\columnwidth]{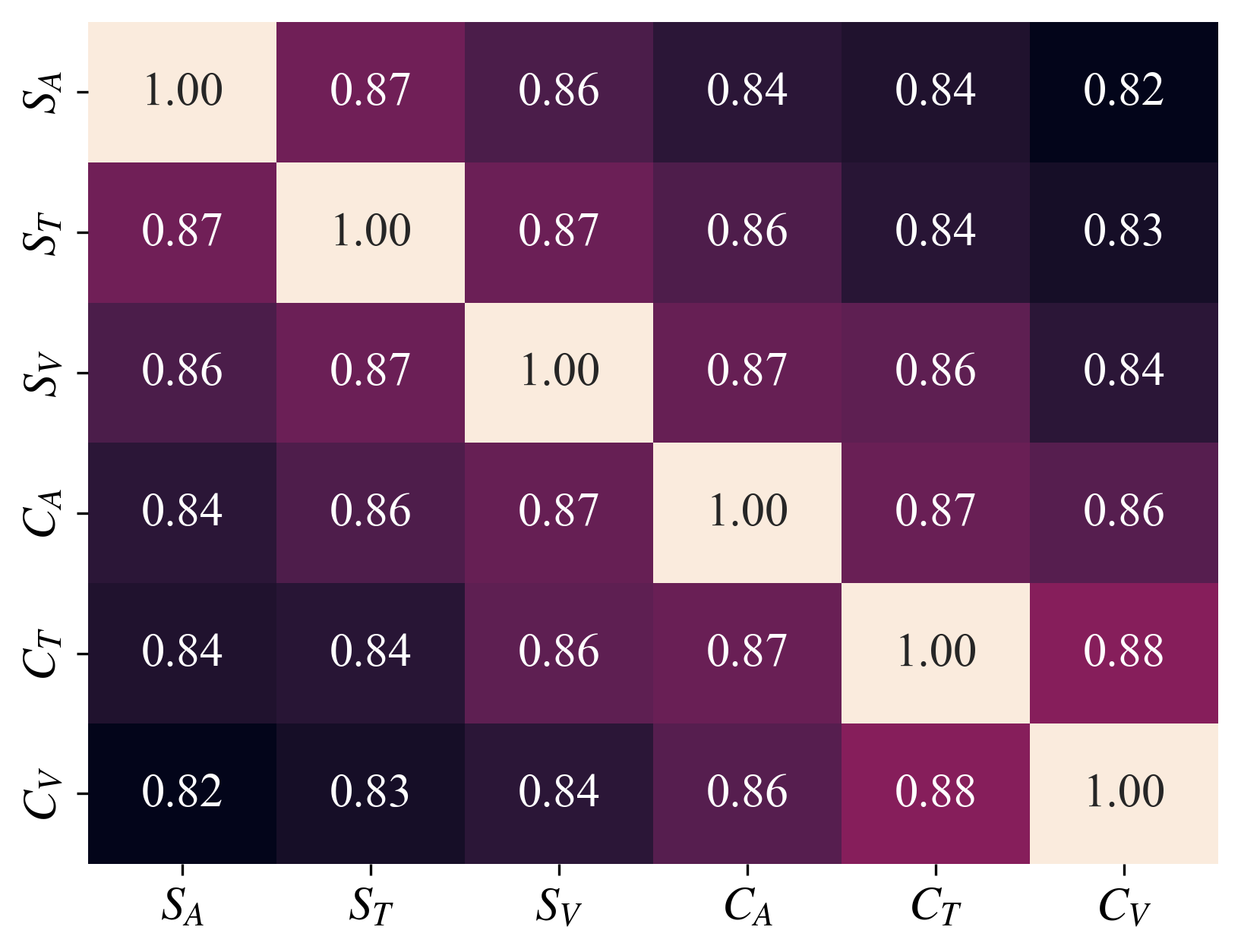}\label{second_av_cos}
 }
 \caption{Visualization of cosine similarity of representations on IEMOCAP dataset.}\label{cosine_simi}
\end{figure*}

\subsubsection{Analysis of Experiments on Noisy Datasets.}
Here we introduce controlled noise to simulate realistic data collection challenges. Specifically, we add gaussian noise of varying intensities \cite{gao2024embracing} to the existing modality representations across both training stages, as well as during validation and testing. Table \ref{EX_NOISY} reports the average performance across seven missing modality scenarios. We observe a clear performance degradation as noise intensity increases. In IEMOCAP, TMDC’s accuracy drops significantly from 73.8/73.6 to 37.1/32.2 when noise intensity reaches 20, while in MOSEI, the impact is less pronounced, with performance declining from 81.2/80.8 to 66.6/63.9. Despite the degradation, TMDC consistently outperforms MoMKE at all noise levels, particularly at a noise intensity of 10, where it surpasses MoMKE by an average of 10 points across all datasets. 
\begin{table}[H]
\centering
\begin{tabular}{cccc}
\hline
 & MOSI & MOSEI & IEMOCAP \\
\hline
\hline
TMDC & \textbf{77.64/77.35} & \textbf{81.22/80.76} & \textbf{73.77/73.64} \\
w/o IMD & 74.48/74.34 & 79.76/79.36 & 72.59/71.76 \\
w/o IMC & 74.17/73.97 & 79.74/79.41 & 66.05/65.94 \\
w/o MCD & 75.67/75.26 & 80.03/79.52 & 72.41/71.99 \\
w/o MSD & 74.76/74.16 & 80.03/79.92 & 71.68/71.24 \\
\hline
\end{tabular}
\caption{Ablation studies on different datasets. The averaged results of all seven conditions are reported. Best performance is bold.}\label{EX_ablas}
\end{table}

\begin{table}[H]
\centering
\begin{tabular}{c|ccc}
\hline
\multirow{2.5}{*}{Methods} & \multicolumn{3}{c}{\textbf{Gaussian Noise}} \\
\cline{2-4}
& $\epsilon=5$ & $\epsilon=10$ & $\epsilon=20$ \\
\hline
\hline
\multicolumn{4}{c}{\textbf{Results on Noisy MOSI}} \\ 
\hline
MoMKE & 55.9/55.8 & 53.9/54.1 & 52.7/52.9 \\
TMDC & \textbf{68.4/67.8} & \textbf{60.8/60.2} & \textbf{53.0/53.0} \\
\hline
\hline
\multicolumn{4}{c}{\textbf{Results on Noisy MOSEI}} \\ 
\hline
MoMKE & 71.0/69.3 & 61.2/58.4 & 59.4/55.4 \\
TMDC & \textbf{74.5/73.7} & \textbf{71.2/70.5} & \textbf{66.3/63.9} \\
\hline
\hline
\multicolumn{4}{c}{\textbf{Results on Noisy IEMOCAP}} \\ 
\hline
MoMKE & 46.1/44.2 & 34.4/30.3 & 31.1/28.3 \\
TMDC & \textbf{60.0/57.9} & \textbf{51.0/48.5} & \textbf{37.1/32.2} \\
\hline
\end{tabular}
\caption{Performance on noisy datasets. The averaged results of all seven conditions are reported.}\label{EX_NOISY}
\end{table}
\subsubsection{Analysis of Training Convergence.}
To evaluate the training stability and convergence of IMD stage, we track the behavior of 12 loss terms over 100 epochs on the MOSI dataset in Figure \ref{training_losses}. The results show a clear correlation between the decline rate of each loss and the performance of its corresponding unimodal representation. As reported in Table \ref{EX_MOSEI}, the text modality outperforms others on MOSI, while the audio modality performs worst. Consistently, text-related loss terms (e.g., $\mathcal{L}_S^T$, $\mathcal{L}_C^T$, $\mathcal{L}_{Spe}^T$, $\mathcal{L}_{Com}^T$) converge more rapidly than those of the audio modality, with the video modality falling in between. We also observe that the VIB-related losses are slightly higher than the interaction-based ones, possibly due to inherent noise in the data. Nonetheless, all losses consistently decrease over time, confirming the effectiveness and stability of the optimization process.

\subsubsection{Analysis of Representations' Relations.}
TMDC learns both modality-specific and modality-invariant representations during training. To analyze their relationships under different settings, we visualize average cosine similarities on the IEMOCAP test set in Figure \ref{cosine_simi}, where $S_m = \hat{X}_{Spe}^m$ and $C_m = \hat{X}_{Com}^m$. We consider four cases: (a) the IMD stage, (b) the IMC stage with all modalities, (c) only audio available, and (d) audio and video available. 

Across most settings, $S_m$ are consistently more similar to each other than to $C_m$, suggesting they capture complementary information. This relationship remains stable under different missing conditions, demonstrating the robustness of the representations. We also observe that similarity scores decrease as more modalities are available (0.87–0.92 with only audio; 0.82–0.88 with audio and video; 0.68–0.77 with all). This indicates that richer modality input encourages more diverse representation learning. Notably, when only audio is present, the inferred text and video representations differ from the audio, showing that the IMC module generates distinct and informative approximations for missing modalities.


\section{Conclusion}
This paper presents TMDC, a Two-stage Modality Denoising and Complementation framework designed to address the challenges of noisy and missing modalities in MSA. TMDC operates in two stages: the first stage reduces noise by learning both modality-specific and modality-invariant features from complete data; the second stage enhances representations by leveraging available modalities to complement the missing ones during training on incomplete data. Extensive experiments on three benchmark datasets under seven missing modality scenarios and high-noise conditions demonstrate that TMDC consistently outperforms existing methods. While the framework achieves strong performance, it introduces some redundancy in the shared representations. Future work will aim to reduce this redundancy to further improve efficiency.

\section*{Acknowledgments}
This work was supported by Sichuan Science and Technology Program (Grant No.2024YFG0006), the National Natural Science Foundation of China (Grant No.U24A20250), and the Fundamental Research Funds for the Central Universities (No.ZYGX2024Z005).

\bibliography{aaai2026}

\clearpage
\appendix
\section{Supplementary Material}

\subsection{Additional Implementation Details}
Consistent with prior work \cite{ xu2024leveraging, guo2024multimodal}, we evaluate the performance of TMDC under fixed missing modality scenarios, where a specific modality is absent during training, validation, and testing. The missing modality is replaced with an all-zero vector. For example, in Tables \ref{EX_MOSEI} and \ref{EX_ABLA}, `A' indicates that only the audio modality is available. It is important to note that, following previous studies\cite{wang2024incomplete,xu2024leveraging,wang2023distribution}, modality missingness is only applied in the IMC stage of TMDC. The IMD stage is trained using the complete dataset. All experiments are implemented in PyTorch and conducted on a GTX 3090 GPU with CUDA 11.5. We use Torch version 1.12.1 for model training and Adam optimizer \cite{kingma2014adam} across all datasets. Due to variations in datasets and modalities, we define a set of default hyperparameters for each dataset: MOSI: $D$ = 256, learning rate = 0.0001, batch size = 32, dropout rate = 0.5, $\beta$ = 0.01, IMD Stage: 80 epochs, IMC Stage: 100 epochs. MOSEI: $D$ = 256, learning rate = 0.0001, batch size = 64, dropout rate = 0.6, $\beta$ = 0.01, IMD Stage: 50 epochs, IMC Stage: 100 epochs. IEMOCAP: $D$ = 256, learning rate = 0.0001, batch size = 16, dropout rate = 0.5, $\beta$ = 0.01, IMD Stage: 50 epochs, IMC Stage: 100 epochs. 

\subsection{Additional Results}
\subsubsection{Detailed Ablation Studies.}
Although TMDC demonstrates strong performance across datasets, it remains unclear how each module and stage contributes to its effectiveness. To investigate this, we conduct an ablation study by evaluating four TMDC variants: (1) `w/o IMD':  Excludes the Intra-Modality Denoising stage, meaning TMDC is not trained on complete datasets before transitioning to the IMC stage. (2) `w/o IMC': Removes Inter-Modality Complementation, where representations from available modalities are directly concatenated without additional compensation after the IMD stage. (3) `w/o MCD': Excludes the Modality-Common Denoising Module in both stages, preventing any compensation for missing information. (4) `w/o MSD': Removes the Modality-Specific Denoising Module in both stages, relying only on shared modality-invariant information for predictions.

Table \ref{EX_ABLA} presents the results of TMDC and its four variants across multiple missing-modality scenarios. We observe that `w/o IMC' performs particularly poorly when only a single modality is available, such as `A' and `V' in MOSI or `A' in MOSEI. This suggests that losing a dominant modality (e.g., `T' in MOSI and MOSEI) severely impacts performance. Overall, removing both denoising modules leads to performance degradation, with `w/o MSD' having a more pronounced impact than `w/o MCD'. This may be because modality-specific information is more discriminative than modality-invariant information, as the latter can be inferred from other modalities, whereas the former cannot. Consequently, losing modality-specific information results in a greater drop in performance.

\begin{table*}[t]
\centering
\begin{tabular}{@{}cccccccccc@{}}
\hline
\multirow{2.5}{*}{Configs} & \multicolumn{7}{c}{\textbf{Modalities}} \\
\cmidrule(r){2-8}  
& A & T & V & A,V & A,T & T,V & T,A,V \\
\hline
\hline
\multicolumn{8}{c}{\textbf{Ablation Results on MOSI}} \\ 
\hline
w/o IMD & 51.07/50.95 & 86.12/85.88 & 61.74/61.15 & 62.20/62.42 & 86.13/85.98 & 86.89/86.85 & 87.20/87.15 \\
w/o IMC & 52.13/51.19 & 86.13/86.06 & 60.37/59.81 & 60.67/60.86 & 85.82/85.85 & 86.89/86.88 & 87.20/87.15 \\
w/o MCD & 54.42/52.59 & 86.43/86.35 & 63.26/63.29 & 64.63/63.70 & 86.28/86.25 & 87.35/87.32 & 87.35/87.33 \\
w/o MSD & 47.56/43.34 & 87.02/87.04 & 62.35/62.30 & 64.02/64.23 & 87.20/87.20 & 87.65/87.60 & 87.50/87.44 \\
TMDC & \textbf{62.35/60.24} & \textbf{87.35/87.27} & \textbf{64.63/64.82} & \textbf{65.40/65.60} & \textbf{87.50/87.45} & \textbf{87.96/87.87} & \textbf{88.26/88.19} \\
\hline
\hline
\multicolumn{8}{c}{\textbf{Ablation Results on MOSEI}} \\ 
\hline
w/o IMD & 72.43/70.48 & 86.32/86.22 & 67.03/66.69 & 71.46/71.25 & 86.76/86.77 & 87.12/87.01 & 87.23/87.13 \\
w/o IMC & 70.61/70.38 & 86.71/86.57 & 69.32/68.20 & 70.64/69.90 & 86.87/86.83 & 86.96/86.93 & 87.09/87.09 \\
w/o MCD & 72.67/71.20 & 86.05/85.98 & 67.58/66.71 & 72.40/71.30 & 86.90/86.85 & 87.29/87.27 & 87.34/87.36 \\
w/o MSD & 72.65/71.45 & 86.68/86.66 & 68.55/67.78 & 72.81/72.12 & 87.04/86.91 & 87.29/87.25 & 87.31/87.28 \\
TMDC & \textbf{73.64/72.23} & \textbf{86.87/86.82} & \textbf{71.60/70.65} & \textbf{74.13/73.41} & \textbf{87.15/87.14} & \textbf{87.48/87.43} & \textbf{87.67/87.62} \\
\hline
\hline
\multicolumn{8}{c}{\textbf{Ablation Results on IEMOCAP}} \\ 
\hline
w/o IMD & 69.60/69.60 & 75.57/75.27 & 57.18/52.78 & 69.01/68.71 & 79.41/79.06 & 77.76/77.53 & 79.62/79.36 \\
w/o IMC & 61.20/62.81 & 65.92/67.17 & 53.22/49.35 & 62.82/61.14 & 73.27/74.71 & 71.07/70.81 & 74.85/75.59 \\
w/o MCD & 68.04/69.39 & 76.77/76.72 & 55.01/51.45 & 69.50/69.33 & 79.68/79.44 & 78.14/77.78 & 79.71/79.84 \\
w/o MSD & 67.58/69.90 & 75.01/74.77 & 53.64/49.61 & 68.28/67.84 & 79.69/79.36 & 77.86/77.44 & 79.71/79.78 \\
TMDC & \textbf{70.45/71.40} & \textbf{77.88/78.44} & \textbf{59.18/55.33} & \textbf{70.21/69.91} & \textbf{79.99/81.45} & \textbf{78.20/78.29} & \textbf{80.48/80.69} \\
\hline
\end{tabular}
\caption{Ablation studies on MOSI, MOSEI, and IEMOCAP datasets.Best performance is bold.}\label{EX_ABLA}
\end{table*}

\begin{table}[]
\centering
\begin{tabular}{ccc}
\hline
Model & Parameters & Training Time  \\
\hline
MPLMM & 128M & 3,433s \\
DiCMoR & 113M & 12,792s \\
IMDer & 122M & 175,138s \\
MoMKE & 5.4M & 39s \\
TMDC & 6.1M & 35s \\
\hline
\end{tabular}
\caption{Computation overhead comparison on MOSI.}
\label{tab:com_overhead}
\end{table}

\begin{table}[]
\centering
\begin{tabular}{cccc}
\hline
Dataset & ACC & F1 \\
\hline
MOSI & 77.64$\pm$0.08 & 77.35$\pm$0.08 \\
MOSEI & 81.22$\pm$0.01 & 80.76$\pm$0.08 \\
\hline
\end{tabular}
\caption{Robustness of TMDC with 5 different random seeds (95\%confidence intervals). Averaged `ACC/F1' of all missing scenarios is reported.}
\label{tab:stas}
\end{table}

\subsubsection{Detailed Performance on Noisy Datasets.} Table \ref{EX_NOISY_ALL} presents the performance of different models under various noise conditions across the MOSI, MOSEI, and IEMOCAP datasets. Overall, TMDC consistently outperforms other methods in most settings. Although its performance slightly lags in certain cases where only the audio (A) or visual (V) modality is available, TMDC demonstrates greater robustness and stability in more complex scenarios. For example, under multiple noisy conditions on the MOSEI and IEMOCAP datasets, TMDC achieves notably better results compared to other baselines.

\begin{table*}[t]
\centering
\begin{tabular}{@{}ccccccccccc}
\hline
\multirow{2.5}{*}{Models} & \multirow{2.5}{*}{$\sigma$} & \multicolumn{7}{c}{\textbf{Modalities}} \\
\cmidrule(r){3-9}
&  & A & T & V & A,V & A,T & T,V & T,A,V \\
\hline
\hline
\multicolumn{9}{c}{\textbf{Results on MOSI}} \\ 
\hline
DiCMoR & 5 & \textbf{53.24}/47.90 & 70.58/70.51 & 56.02/49.02 & 55.09/54.72 & 72.87/72.99 & 69.97/69.75 & 72.22/71.56 \\
MoMKE & 5 & 52.90/\textbf{52.25} & 58.69/58.90 & 52.13/51.83 & 52.29/51.79 & 59.49/59.41 & 57.77/57.93 & 57.93/58.15 \\
TMDC & 5 & 46.80/43.06 & \textbf{77.90/77.84} & \textbf{59.15/59.37} & \textbf{59.91/59.95} & \textbf{77.90/77.84} & \textbf{78.96/78.98} & \textbf{77.90/77.56} \\
\hline
DiCMoR & 10 & 50.46/45.91 & 59.72/57.88 & 51.85/47.91 & \textbf{54.63}/50.62 & 61.57/58.04 & 63.43/59.81 & 60.65/58.66 \\
MoMKE & 10 & \textbf{51.68/51.69} & 54.42/54.69 & \textbf{53.96/54.18} & 53.81/54.05 & 54.27/54.33 & 54.57/54.78 & 54.88/54.96 \\
TMDC & 10 & 46.80/43.62 & \textbf{70.12/70.21} & 49.70/47.90 & 54.42/\textbf{54.67} & \textbf{69.51/69.69} & \textbf{67.99/68.18} & \textbf{66.77/66.97} \\
\hline
DiCMoR & 20 & 48.32/44.69 & 50.00/47.67 & 47.87/45.36 & 42.68/38.71 & 50.46/46.80 & 50.15/48.34 & 50.61/48.31 \\
MoMKE & 20 & \textbf{52.44/52.61} & 54.27/54.51 & \textbf{53.35/53.64} & 51.22/51.36 & 52.13/52.42 & \textbf{53.35/53.60} & 51.83/51.97 \\
TMDC & 20 & 50.46/50.26 & \textbf{54.57/54.83} & 52.13/52.39 & \textbf{53.81/52.63} & \textbf{54.57/54.82} & 52.90/52.86 & \textbf{53.05/53.18} \\
\hline
\hline
\multicolumn{9}{c}{\textbf{Results on MOSEI}} \\ 
\hline
MoMKE & 5 & 60.48/55.91 & 77.74/77.67 & 63.87/60.83 & 63.04/59.63 & 77.66/77.10 & 77.11/76.94 & 77.22/76.89 \\
TMDC & 5 & \textbf{62.25/57.97} & \textbf{80.24/80.17} & \textbf{68.99/68.33} & \textbf{68.71/68.74} & \textbf{80.46/80.12} & \textbf{80.63/80.23} & \textbf{80.13/80.03} \\
\hline
MoMKE & 10 & \textbf{60.37}/55.01 & 62.25/60.66 & 61.89/55.25 & 60.76/56.46 & 60.48/59.90 & 61.28/61.44 & 61.20/59.93 \\
TMDC & 10 & 57.68/\textbf{55.18} & \textbf{76.17/76.02} & \textbf{67.58/67.15} & \textbf{66.87/66.56} & \textbf{76.03/75.78} & \textbf{77.19/76.52} & \textbf{76.91/76.20} \\
\hline
MoMKE & 20 & 59.74/55.50 & 60.13/55.45 & 57.65/55.00 & 58.06/54.85 & 58.67/56.52 & 60.26/54.95 & 61.59/55.37 \\
TMDC & 20 & \textbf{60.43/55.89} & \textbf{70.86/70.62} & \textbf{63.10/62.08} & \textbf{65.05/57.05} & \textbf{63.92/61.27} & \textbf{71.02/70.16} & \textbf{69.95/69.90} \\
\hline
\hline
\multicolumn{9}{c}{\textbf{Results on IEMOCAP}} \\
\hline
MoMKE & 5 & 32.75/\textbf{29.01} & 48.99/49.48 & 42.83/37.65 & 42.57/37.75 & 50.28/50.84 & 52.45/51.92 & 52.66/52.45 \\
TMDC & 5 & \textbf{33.40}/25.85 & \textbf{67.86/67.88} & \textbf{48.32/44.35} & \textbf{57.05/52.11} & \textbf{70.37/72.01} & \textbf{71.44/72.09} & \textbf{71.30/71.32} \\
\hline
MoMKE & 10 & 31.47/\textbf{27.67} & 32.71/29.46 & 39.08/31.17 & 36.83/31.77 & 32.15/28.78 & 34.36/31.32 & 34.43/31.65 \\
TMDC & 10 & \textbf{33.18}/25.77 & \textbf{45.64/43.91} & \textbf{43.88/39.52} & \textbf{46.17/41.05} & \textbf{62.64/62.68} & \textbf{63.12/64.07} & \textbf{62.23/62.54} \\
\hline
MoMKE & 20 & 31.11/\textbf{28.27} & \textbf{31.19}/28.03 & \textbf{31.37}/28.35 & 31.16/28.44 & 30.69/28.10 & 31.28/28.34 & 30.98/28.62 \\
TMDC & 20 & \textbf{33.60}/25.78 & 30.11/\textbf{28.51} & 30.76/\textbf{28.95} & \textbf{43.94/37.10} & \textbf{34.57/30.22} & \textbf{43.50/37.68} & \textbf{42.85/37.21} \\
\hline
\end{tabular}
\caption{Performance on noisy datasets. Best performance is bold.}\label{EX_NOISY_ALL}
\end{table*}

\subsubsection{Hyper-parameter Analysis.} 
\begin{figure*}[h]
    \centering
    \subfloat[`A', `V', and `A,V']{
\includegraphics[width=0.48\columnwidth]{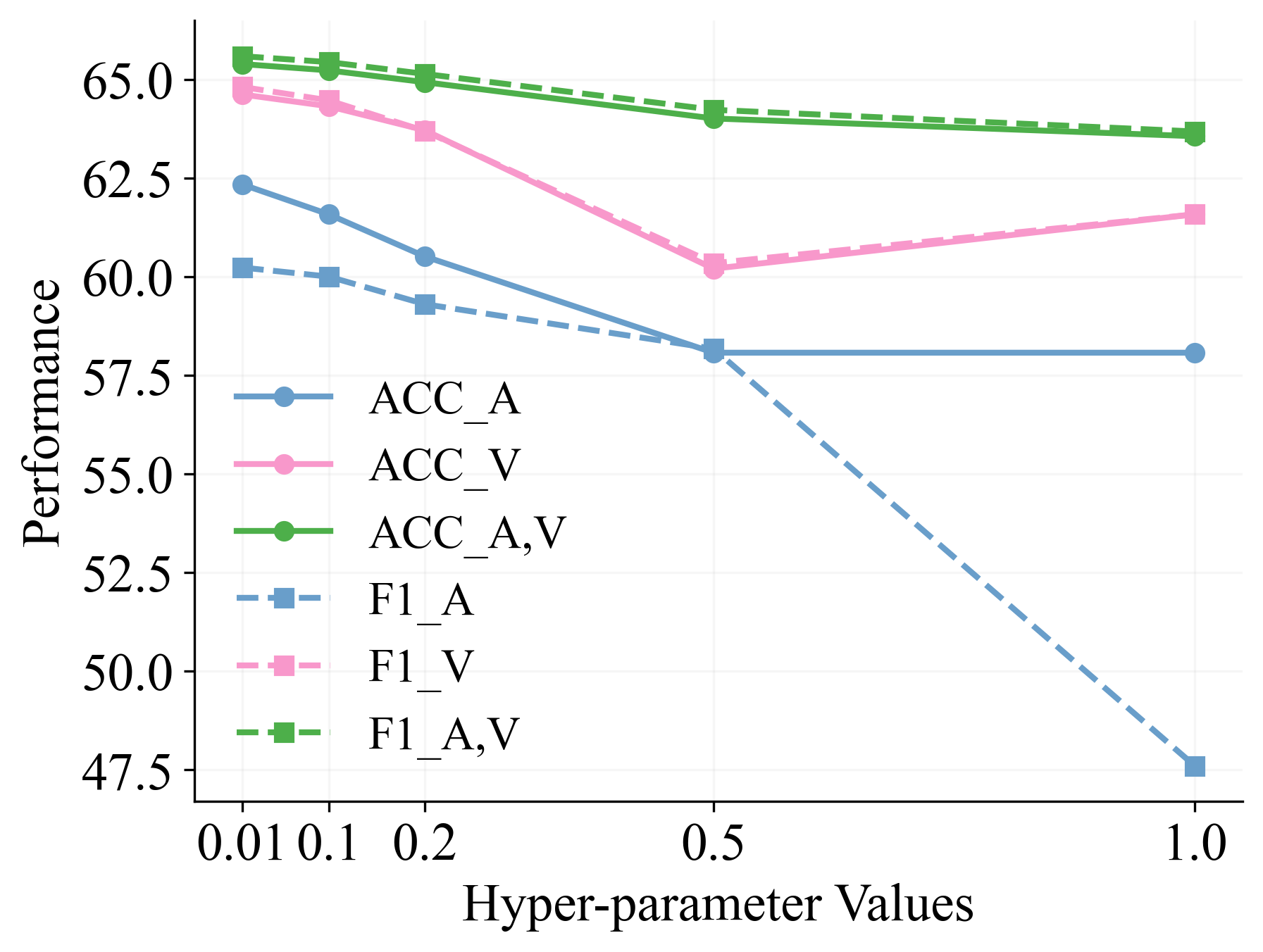}\label{hyper_1}
 }
 \subfloat[`T' and `A,T']{
\includegraphics[width=0.48\columnwidth]{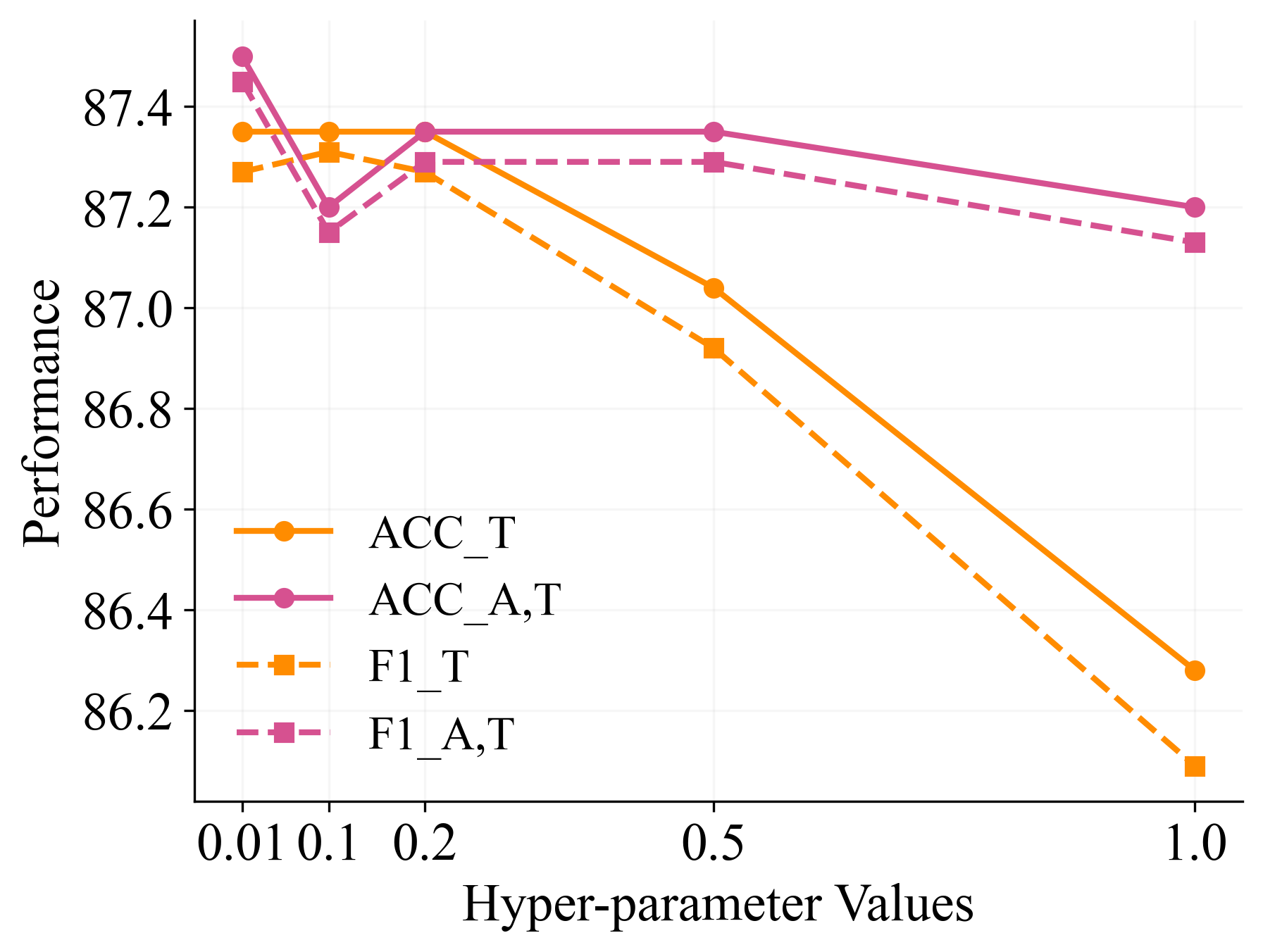}\label{hyper_2}
 }
  \subfloat[`T,V' and `A,T,V']{
\includegraphics[width=0.48\columnwidth]{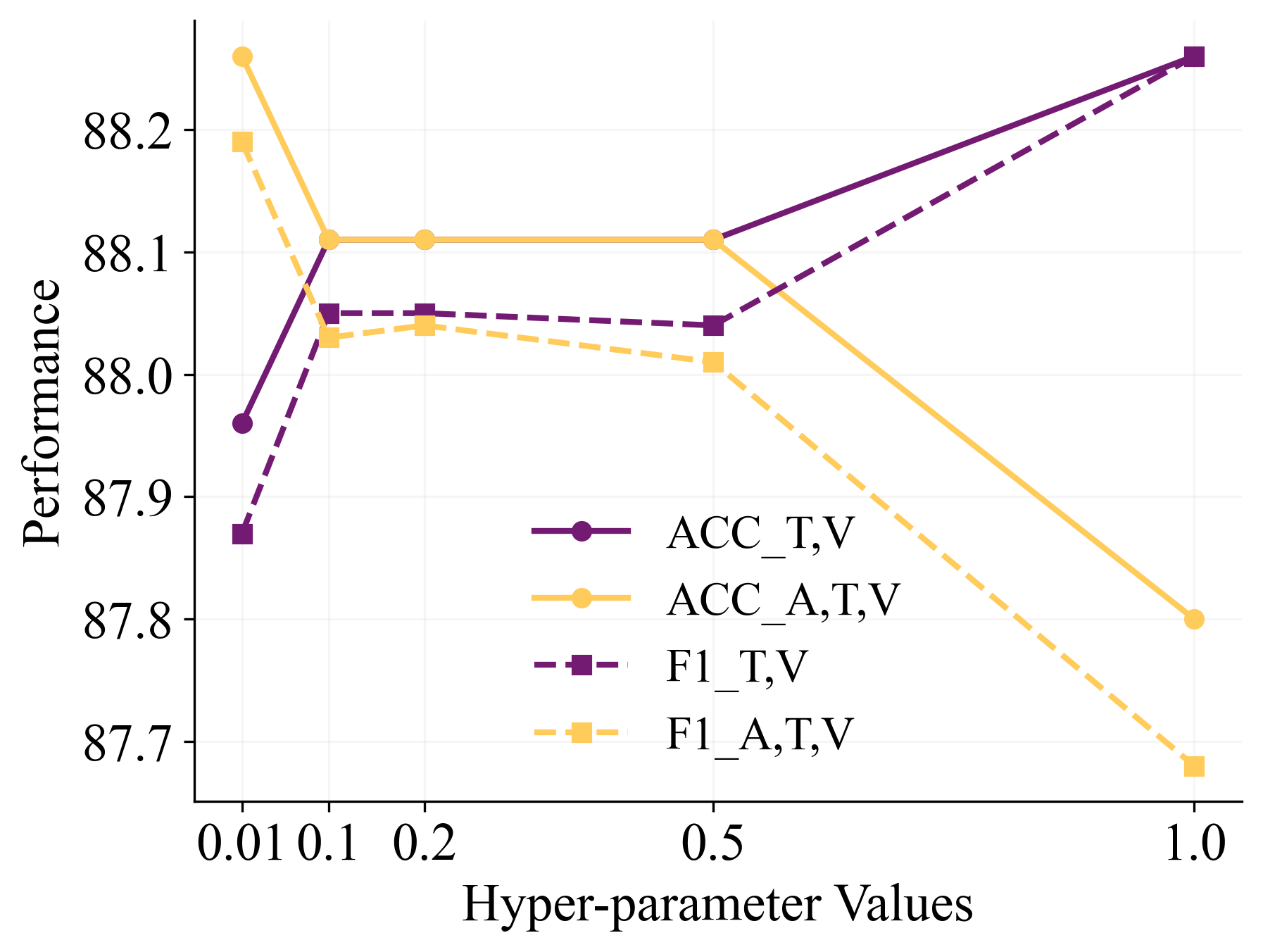}\label{hyper_3}
 }
  \subfloat[AVG]{
\includegraphics[width=0.48\columnwidth]{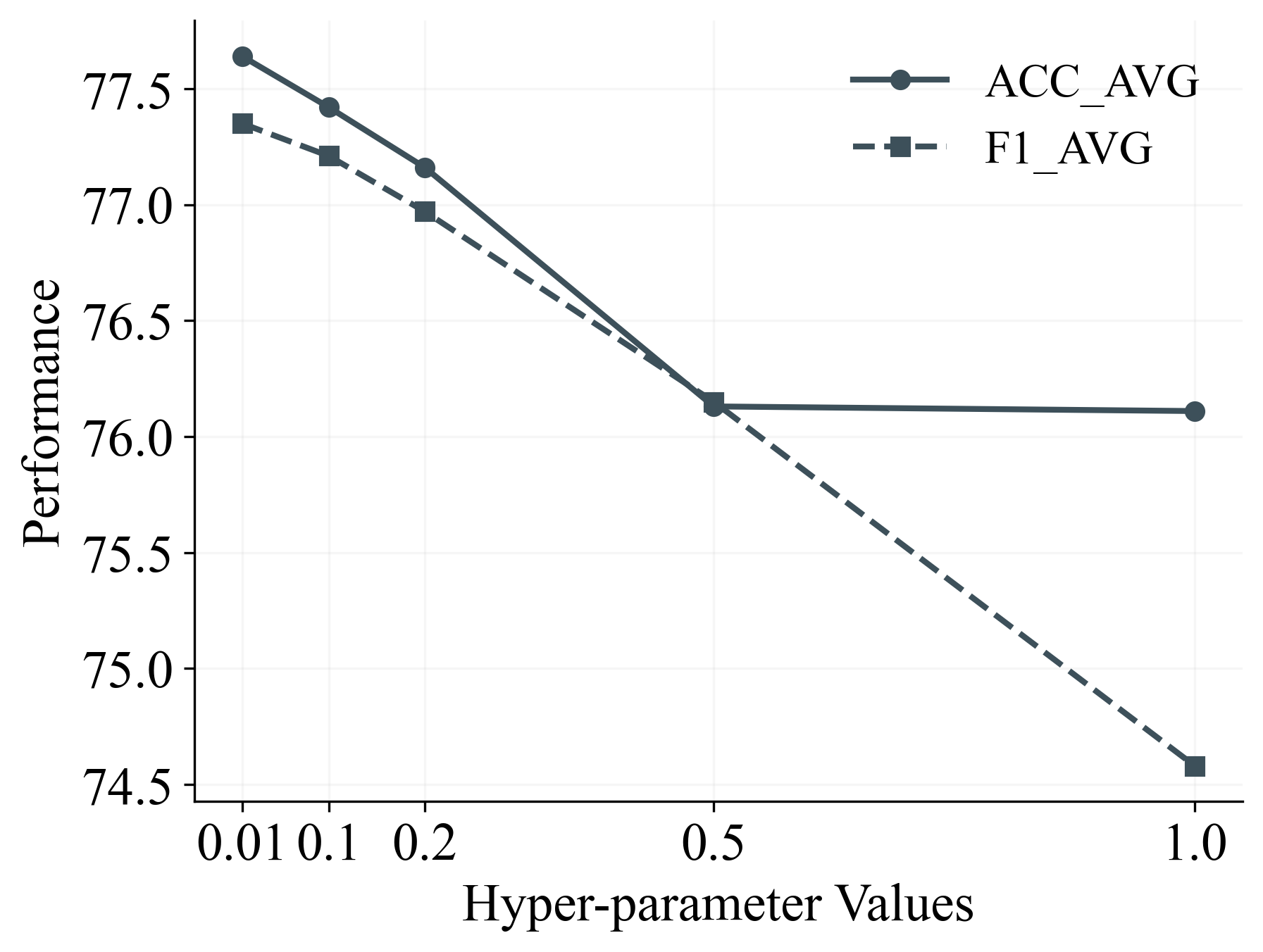}\label{hyper_4}
 }
 \caption{Analysis of hyper-parameter $\beta$ in VIB loss.}\label{hyper_analysis}
\end{figure*}

To further assess TMDC’s robustness to hyper-parameter, Figure \ref{hyper_analysis} analyzes the impact of the VIB loss coefficient $\beta$ under different missing modality scenarios on the MOSI dataset. We observe that the effect of $\beta$ varies across scenarios. In general, larger $\beta$ values—indicating a stronger weight on the VIB loss—tend to degrade performance. However, an exception occurs when only the text (T) and visual (V) modalities are present, where a higher $\beta$ slightly improves results. Overall, TMDC remains stable, with only minor average performance fluctuations as $\beta$ varies between 0.01 and 0.2 (Accuracy: 77.16–77.64).

\subsubsection{Computation Overhead.} We conducted runtime benchmarking on the MOSI dataset using PyTorch on a single RTX 3090 GPU (CUDA 11.5). Results are summarized in Table \ref{tab:com_overhead}. TMDC achieves strong efficiency with only a marginal increase in parameter count compared to MoMKE. It significantly outperforms heavier models like IMDer and DiCMoR in terms of training time, largely due to its lightweight design and the absence of computationally intensive components like diffusion or flow-based modules.

\subsubsection{Robustness of TMDC.} To comprehensively evaluate the robustness of TMDC, Table \ref{tab:stas} reports the standard deviations over five runs. On the MOSI and MOSEI datasets, TMDC achieves 77.64$\pm$0.08 / 77.35$\pm$0.08 and 81.22$\pm$0.01 / 80.76$\pm$0.08, respectively. These results demonstrate that TMDC not only achieves strong performance but also maintains high stability across runs. The consistent improvements over recent state-of-the-art methods, including MPLMM (ACL 2024), IMDer (NeurIPS 2023), and DiCMoR (ICCV 2023), further validate the effectiveness and robustness of our approach.

\subsection{Visualization Analysis}
\begin{figure*}[h]
    \centering
 \subfloat[TMDC]{
\includegraphics[width=0.39\columnwidth]{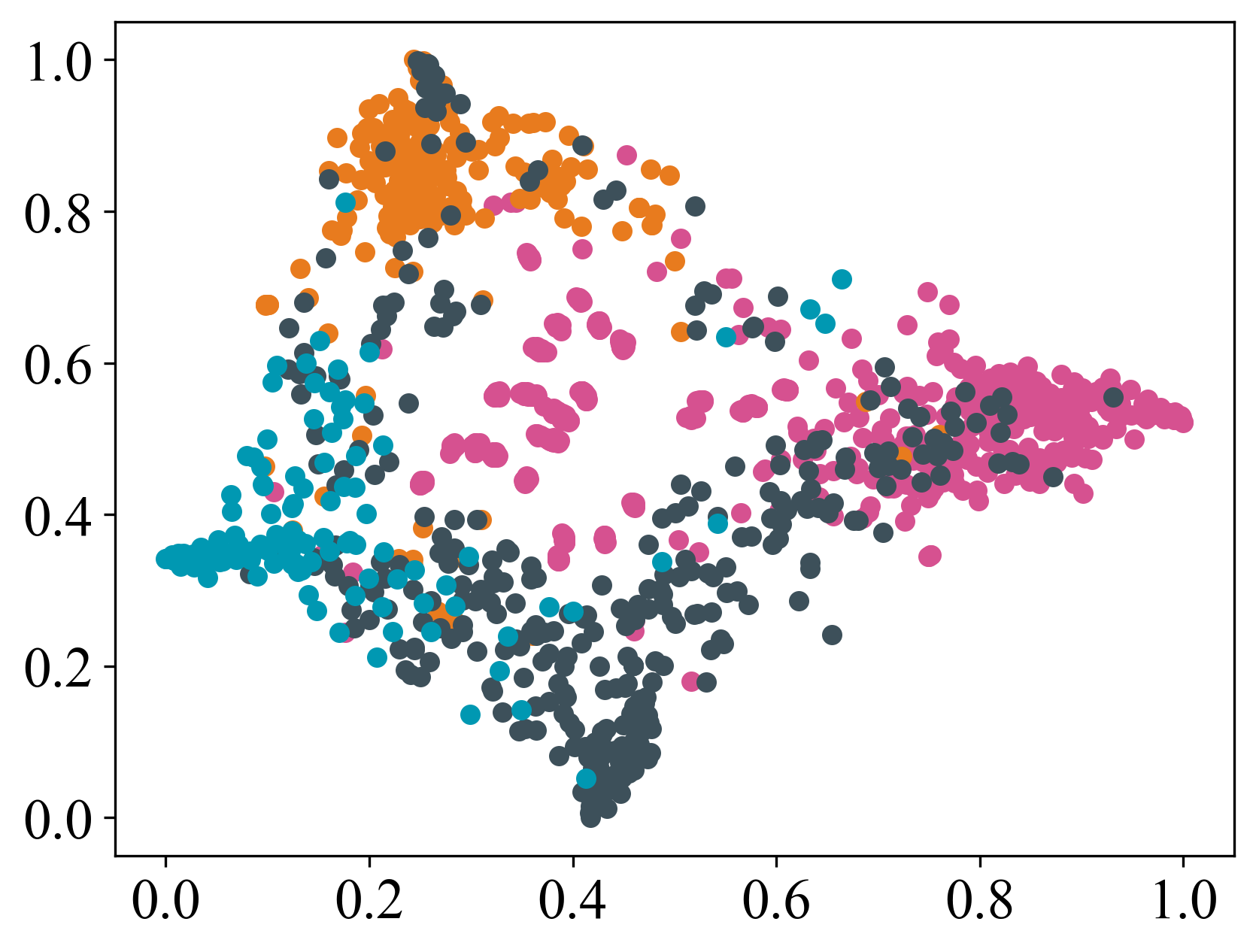}\label{embed_all}
 }
 \subfloat[w/o IMD]{
\includegraphics[width=0.39\columnwidth]{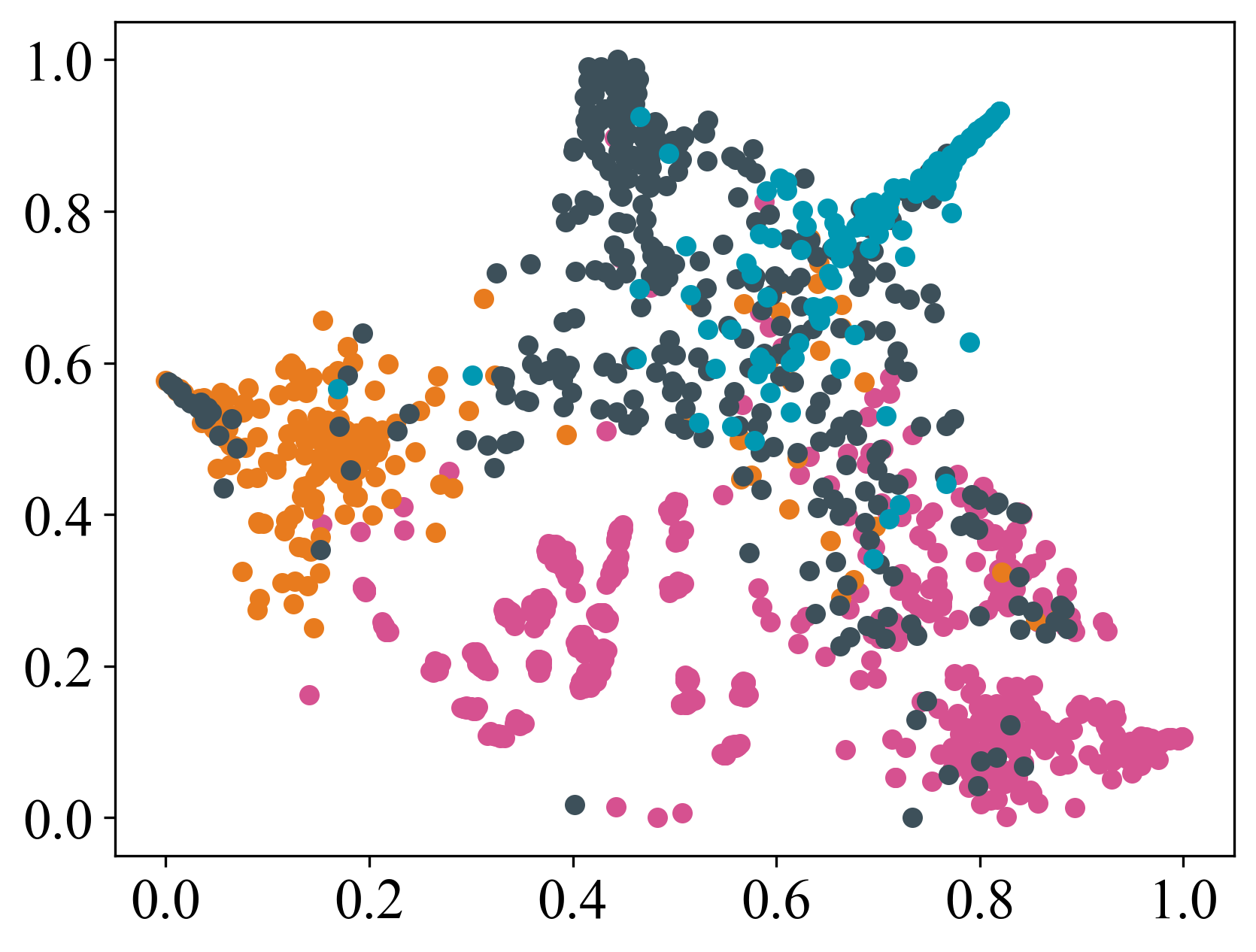}\label{embed_wo_stage1}
 }
 \subfloat[w/o IMC]{
\includegraphics[width=0.39\columnwidth]{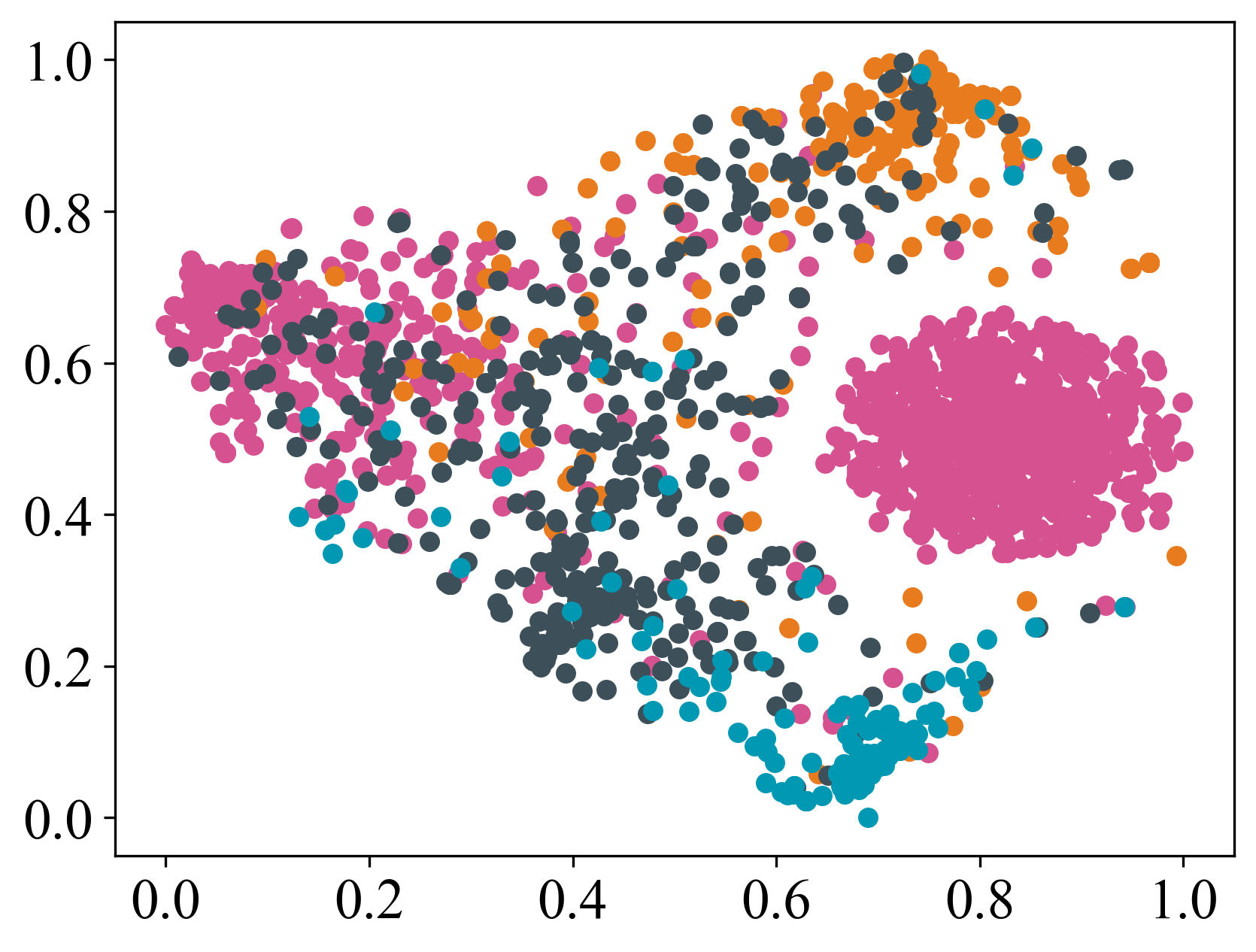}\label{embed_wo_stage2}
 }
\subfloat[w/o MSD]{
\includegraphics[width=0.39\columnwidth]{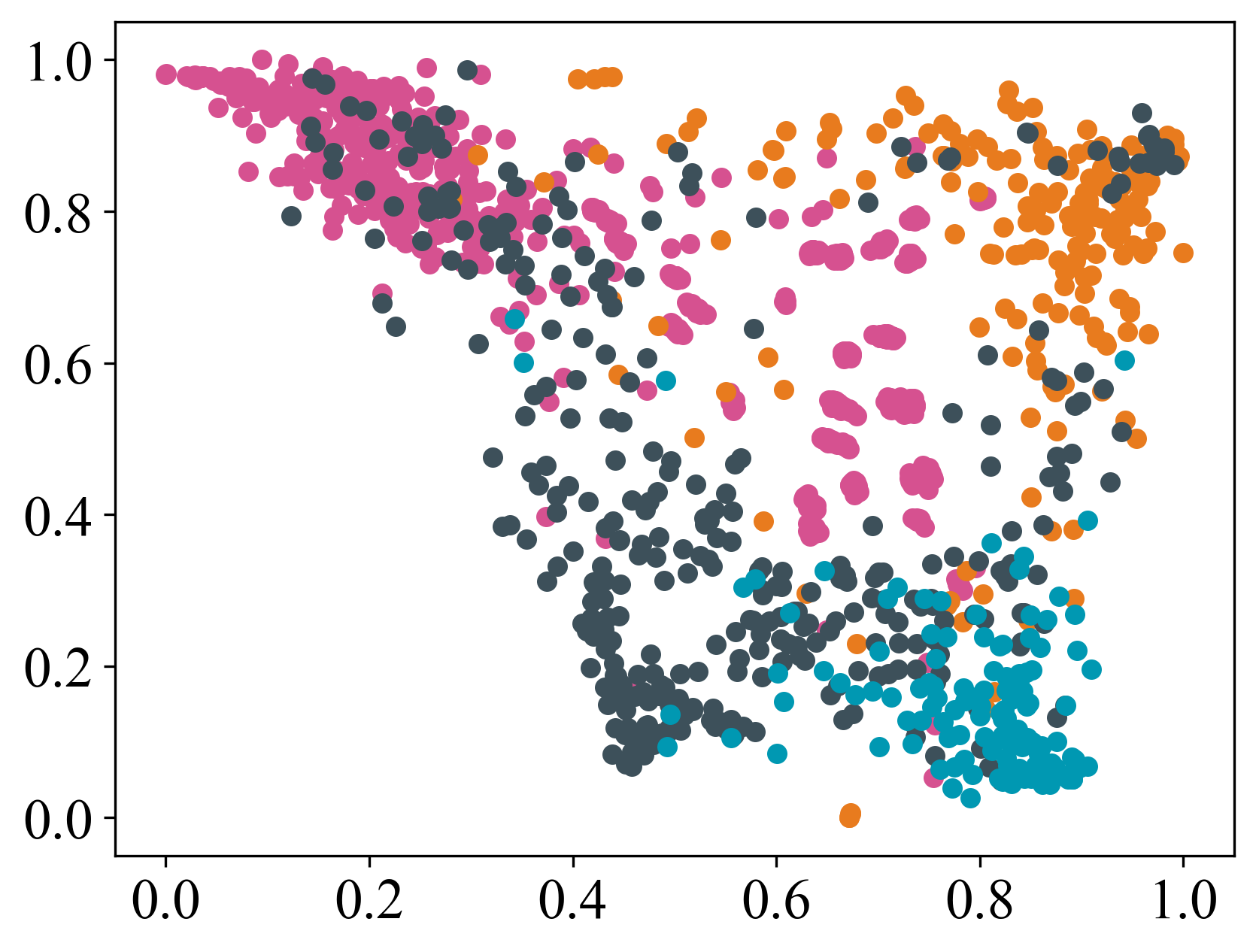}\label{embed_wo_com}
 }
 \subfloat[w/o MCD]{
\includegraphics[width=0.39\columnwidth]{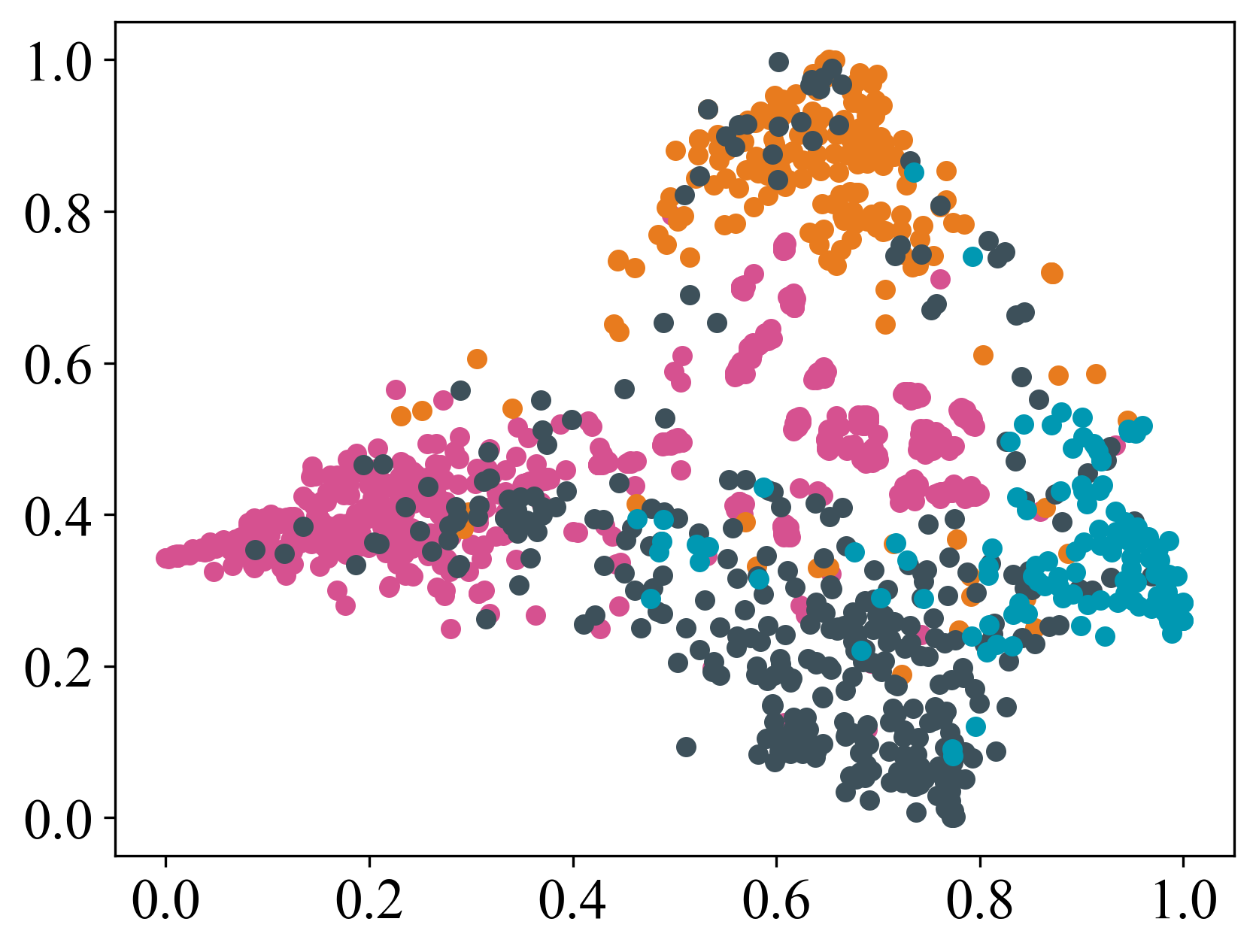}\label{embed_wo_spe}
 }
\caption{t-SNE of the fused representations on IEMOCAP when only text modality is present. We use red, orange, black and blue to represent neutral, happiness, sadness and anger, respectively.}\label{visual_embedds}
\end{figure*}

To further evaluate the effectiveness of TMDC, we visualize the classification performance using t-SNE on the IEMOCAP dataset using only the text modality in the IMC stage. Figure \ref{visual_embedds} compares TMDC with four variant models. In the visualization, neutral, happiness, sadness, and anger are represented in red, orange, black, and blue, respectively.

The results indicate that removing any stage or module degrades classification performance. Specifically, eliminating IMD causes all emotion categories—except anger—to become more dispersed. Removing IMC leads to the neutral category splitting into two distinct clusters, resulting in a sharp accuracy drop. While removing a single module has a less severe impact than removing IMC, the resulting representations are still more scattered compared to the full TMDC model. These findings highlight the importance of each stage and module in TMDC, demonstrating their contributions to maintaining well-structured and distinct emotion representations.

\end{document}